\documentclass[structabstract]{aa}  
\usepackage{graphicx}
\usepackage{txfonts}
\begin{document}
   \title{AKARI observations of ice absorption bands towards edge-on YSOs}

   \subtitle{}

   \author{Y. Aikawa    \inst{1},
              D. Kamuro \inst{1},
              I. Sakon \inst{2},
              Y. Itoh \inst{1},
              H. Terada \inst{3},
              J.A. Noble \inst{4}
              K.M. Pontoppidan \inst{5},
              H.J. Fraser \inst{4},
              M. Tamura \inst{3},
              R. Kandori \inst{3}.
              A. Kawamura \inst{6}
          \and
          M. Ueno\inst{7}
          }

   \institute{Department. of Earth and Planetary Sciences, Kobe University,
              Kobe 657-8501, Japan\\
              \email{aikawa@kobe-u.ac.jp}
         \and
             Department of Astronomy, School of Science, University of Tokyo, 7-3-1
             Hongo, Bunkyo-ku, Tokyo 113-0033, Japan
         \and
             National Astronomical Observatory of Japan, Osawa 2-21-1, Mitaka, Tokyo 181-8588, Japan
         \and
             Department of Physics, Scottish Universities Physics Alliance (SUPA), 
             University of Strathclyde, Glasgow G4 ONG, UK.
         \and
             California Institute of Technology,
             Division of Geological and Planetary Sciences, Pasadena, CA 91125
         \and
             Department of Physics, Nagoya University, Furo-cho, Chikusa-ku, Nagoya 464-8602, Japan
          \and
             Institute of Space and Astronautical Science, Japan Aerospace Exploration Agency, 3-1-1
             Yoshinodai, Sagamihara, Kanagawa 229-8510, Japan }

   \date{Received; accepted}

% \abstract{}{}{}{}{} 
% 5 {} token are mandatory
 
  \abstract
   {Circumstellar disks and envelopes of low-mass young stellar objects (YSOs) contain significant amounts of
ice. Such icy material will evolve to become volatile components of planetary systems, such as comets
in our solar system.}
   {To investigate the composition and evolution of circumstellar ice around low-mass YSOs, we
observed ice absorption bands in the near infrared (NIR) towards eight YSOs ranging from class 0
to class II, among which seven are associated with edge-on disks.}
   {We performed slit-less spectroscopic observations using the grism mode of the Infrared Camera (IRC) on board AKARI,
   which enables us to obtain full NIR spectra from 2.5 $\mu$m to 5 $\mu$m, including the CO$_2$ band
  and the blue wing of the H$_2$O band, which are inaccessible from the ground. We developed
 procedures to carefully process the spectra of targets with nebulosity. The spectra were fitted with polynomial baselines
to derive the absorption spectra. The molecular absorption bands were then fitted with the laboratory database of ice absorption bands, considering the instrumental line profile and the spectral resolution of the grism dispersion element.}
   {Towards the class 0-I sources (L1527, IRC-L1041-2, and IRAS04302), absorption bands of H$_2$O, CO$_2$, CO, and XCN are
   clearly detected.
   Column density ratios of CO$_2$ ice and CO ice relative to H$_2$O ice are  $21-28$ \% and $13-46$ \%, respectively.
   If XCN is OCN$^-$, its column density is as high as $2-6$ \% relative to H$_2$O ice. The HDO ice feature at
   4.1 $\mu$m is tentatively detected towards the class 0-I sources and HV Tau. Non-detections of the CH-stretching mode
  features around 3.5 $\mu$m provide upper limits to the CH$_3$OH abundance of 26 \% (L1527) and 42 \% (IRAS04302)
  relative to H$_2$O.
   We tentatively detect OCS ice absorption towards IRC-L1041-2. Towards class 0-I sources, the detected features
   should mostly originate in
   the cold envelope, while CO gas
   and OCN$^-$ could originate in the region close to the protostar, where there are warm temperatures and UV radiation.
    We detect H$_2$O ice band towards ASR41 and 2MASSJ1628137-243139, which are edge-on class II disks.
    We also detect H$_2$O ice and CO$_2$ ice towards HV Tau, HK Tau, and UY Aur, and tentatively detect
    CO gas features towards HK Tau and UY Aur.}
   {}

   \authorrunning{Aikawa et al.}
   \titlerunning{AKARI observations of ice absorption bands towards edge-on YSOs}
   \maketitle
%
%________________________________________________________________

\section{Introduction}

In molecular clouds, protostellar envelopes, and protoplanetary disks,
significant amounts of oxygen and carbon are in the form of
molecules in ice mantles, such as H$_2$O, CO, CO$_2$, and CH$_3$OH
(e.g. Whittet \cite{whittet93}, Murakawa et al. \cite{murakawa00}, Gibb et al. \cite{gibb04}).
These interstellar ices
are formed by the adsorption of gas-phase molecules onto grain surfaces and/or
the grain-surface reactions of the adsorbed species (e.g. Aikawa et al. \cite{aikawa05}).

Ices in circumstellar disks of low-mass YSOs are of special interest as they contribute to the raw material
of planetary formation. Observations of disk ices, however, are not straightforward
for the following two reasons, and thus remain limited in number. Firstly, ices exist mostly near
the midplane at
$r\ge$ several AU. Observations of disk ices are basically restricted to edge-on disks,
where the ice bands absorb stellar light, scattered
stellar light, and/or thermal emission of warm dust in the disk inner radius
(Pontoppidan et al \cite{crbr05}). The targets are naturally heavily extincted by dust, and thus are faint.
Secondly, it is unclear whether the ice absorption bands, if detected,
originate in the disk or other foreground components (i.e. ambient clouds).
The source CRBR 2422.8-3423 was the first edge-on object towards which detailed ice observations
were performed (Thi et al. \cite{thi02}). Pontoppidan et al. (\cite{crbr05}) concluded, via
detailed analysis and modeling of the object, that only a limited amount ($< 20$ \%) of detected CO ice
may exist in the disk, while up to 50 \% of water and CO$_2$ may originate in the
disk. They also found that the 6.85 $\mu$m band, which is tentatively attributed to NH$_4^+$, has a prominent
red wing. Since this wing is reproduced by the thermal processing in the laboratory, it indicates that along the line of 
sight towards CRBR 2422.8-3423 there is warm ice in the disk.

Honda et al. (\cite{honda09}) succeeded in detecting water ice in a disk around
a Herbig Ae star, which is not
edge-on. They observed scattered light from a disk around HD142527 at multiple wavelength
bands around $\sim 3$ $\mu$m. This method can be a powerful tool for investigating the spatial
distribution of ices in disks. However, it would currently be difficult to detect ice features other than the
3 $\mu$m water band via scattered light.

In the present paper, we report near infrared (NIR) ($2.5-5$ $\mu$m) observations of
edge-on disks using the grism mode of the IRC on board AKARI.
We selected seven edge-on young stellar objects (YSOs); three are considered to be in a transient state of
class 0-I, and four are class II (see \S 2 for more detailed descriptions).
Although the absorption by the protostellar envelope may overwhelm the absorption by the disk in class 0-I YSOs,
envelope material along our line of sight (perpendicular to the rotation axis and outflow) is likely
to end up in the midplane region of the forming disk, and ices with a sublimation temperature of
$> 50$ K (i.e. CO$_2$ and H$_2$O) would survive the accretion shock onto the disk (Visser et al.
\cite{visser09}). 
%Variations of ice composition in the evolutionary sequence
%would help us to constrain the composition of disk ice.

Ice observations towards low-mass YSOs have been performed using {\it Spitzer Space Telescope (SST)}
and various ground-based telescopes such as the Very Large Telescope (VLT) and Subaru.
Pontoppidan et al. (\cite{pontoppidan2003}) observed the
4.67 $\mu$m CO ice band, which is observable from the ground, towards 39 YSOs. Using {\it SST},
Furlan et al.
(\cite{furlan08}) and Pontoppidan et al. (\cite{pontoppidan2008}) observed the strong CO$_2$ ice band at
$15.3$ $\mu$m towards class I sources. Observations with {\it SST} also detected absorption bands of CH$_3$OH, HCOOH,
H$_2$CO, and CH$_4$ (Boogert et al. \cite{boogert08}, Zasowski et al. \cite{zaowski09}, {\rm \"{O}}berg et al.
\cite{oberg_ch4}).
Since ground-based observations are restricted to atmospheric windows, and since {\it SST} is
restricted to the mid infrared ($5-36$ $\mu$m), NIR observations via AKARI
are important to comprehend the {\it full} spectra of ices in circumstellar material.
AKARI has a high enough sensitivity for the spectroscopic observation of low-mass YSOs;
the sensitivity of the AKARI grism mode at 3 $\mu$m is 120 $\mu$Jy (1 $\sigma$, 10min), which
is almost comparable to the K-band sensitivity of the spectroscopic observation at 8-m ground-based
telescopes (VLT and Subaru). While the L-band sensitivity on the ground is $\sim 5$ mag or 100 times
worse than at K-band, the sensitivity of the AKARI grism mode does not vary significantly at $2.5- 
4$ $\mu$m, and only gradually worsens at longer wavelengths, by up to a factor of 3-4 at 5 $\mu$m.

\section{Observations}
Observations were performed using the IRC (Onaka et al \cite{onaka07}, Ohyama et al. \cite{ohyama07})
on board the AKARI satellite (Murakami et al \cite{murakami07}) from September 2006 to
March 2007. We used the Astronomical Observational Template 04 (AOT04), designed
for spectroscopy. The AOT04 replaces the imaging
filters with transmission-type dispersers on the filter wheels to
take near- and mid-infrared spectra. The field of view (FOV) consists of a 1'$\times$ 1' aperture
(covered by 40 $\times$ 40 pixels in the imaging mode) (Figure \ref{N3}a)
and a 10'$\times$ 10' aperture. The targets are set
at the center of the 1'$\times$ 1' aperture (without a slit),
where only the NIR channel is available. 
The resolving power of the IRC grism mode is 0.00965 $\mu$m/pixel, and the FWHM of the point spread function (PSF)
is $\sim 2.9$ pixels in the NIR. The spectral resolution is thus $\sim 100$, which is sufficient to detect the ice absorption bands.
In each observation, the following data are taken in a period of 600 s:
two dark frames, one image frame with the N3 filter ($2.7-3.8 \mu$m),
and eight slitless spectral images.

Table \ref{targets} lists our targets, and Figure \ref{N3} (a) shows
the N3 images of our targets. The stellar light is dispersed from right (short wavelengths) to left (long wavelengths).
The position angle of the FOV is fixed; the dispersion direction is almost aligned with the ecliptic latitude.
The nebulosity of our targets extends almost perpendicularly to the dispersion direction.
As an example, the spectral images of L1527 and IRAS04302 are shown in Figure \ref{N3} (b).

Both IRAS04302 and L1527 (IRAS 04368+2557) are edge-on YSOs
in a transient state from class 0 to class I (Andr\'{e} et al. \cite{AWB2000}).
The YSO L1527 is of special interest from
a chemical point of view; it is one of two YSOs where high abundances
of carbon-chain species are observed. Carbon chains are considered to be
formed by warm carbon-chain chemistry (WCCC): gas-phase reactions
of sublimated CH$_4$ (Sakai et al. \cite{nami08}, Aikawa et al. \cite{aikawa08}).
The large organic species detected towards several low-mass YSOs
(e.g. IRAS16293) are not, however, detected towards L1527
(Sakai et al. \cite{nami08}). Since the large organic species
are formed from CH$_3$OH, while carbon chains are formed from CH$_4$,
it is important to investigate ice bands, including the C-H bands, towards L1527, to
determine whether its ice composition significantly differs from those of other low-mass YSOs.

The YSO IRC-L1041-2 was initially in our target list of
background stars (i.e. field stars behind molecular clouds) (Noble et al. in prep).
The N3 image shows a faint object with nebulosity, which is similar to L1527.
The {\it SST} observations also detected a YSO with nebulosity (T. Bourke private communication).
The presence of a dark  dust lane indicates that the YSO has a high inclination, close to edge-on.

The YSOs 2MASS J1628137 and ASR41 are edge-on class II objects. 2MASS J1628137 is accompanied
by a tenuous ($4\times 10^{-4}$ M$_{\odot}$) envelope (Grosso et al. {\cite{grosso03}). 
Hodapp et al. (\cite{hodapp04})  performed detailed modeling of the scattering and radiative
transfer of ASR41 and concluded that the K-band image can be well reproduced
by a protoplanetary disk embedded in a low-density cloud.

HK Tau B and HV Tau C are edge-on class II objects, in the HK Tau binary and HV Tau triplet systems,
respectively. We initially aimed to obtain spectra of these edge-on objects, but the spectra
were found to be contaminated and dominated by the bright primaries, because the binary separations
are small (see \S 4.1 and \S 5). The ice absorption features towards HK Tau B and HV Tau C might still
be observable, although the derivation of ice column densities is not straightforward.
For a comparison with disks that are not edge-on, we also observed UY Aur, which is a binary system with a
separation of $0.9"$ accompanied by a circumbinary disk. The inclination of the disk is $\sim 42 ^{\circ}$
(Hioki et al. \cite{hioki07}).

\begin{table*}
\caption{Target list}
\label{targets}
\centering
\begin{tabular}{c c c l}
\hline\hline 
Name & coordinates (J2000)$^{\mathrm{a}}$ 
 & Class & note\\    % table heading 
\hline                        % inserts single horizontal line
L1527(IRAS 04368+2557) & 04:39:53.6 +26:03:05.5 & 0-I & WCCC,  $i\sim 85^{\circ}$ $^{\mathrm{b}}$\\
IRC-L1041-2 & 20:37:21, +57:44:13 & 0-I $^{\mathrm{b}}$& nearly edge-on$^{\mathrm{c}}$\\
IRAS04302+2247   & 04:33:16.45, +22:53:20.7 & 0-I & $i\sim 90^{\circ}$ $^{\mathrm{d}}$ \\
ASR 41      & 03:28:51.291, +31:17:39.79 &  II &  $i\sim 80^{\circ}$ $^{\mathrm{e}}$\\
2MASSJ1628137-243139 & 16:28:13.7, -24:31:39.00 & II &   $i\sim 86^{\circ}$ $^{\mathrm{f}}$ \\
HV Tau     & 04:38:35.280, +26:10:39.88 & II  & multiple system ($i\sim 84^{\circ}$ $^{\mathrm{g}}$) \\
HK Tau     & 04:31:50.900, +24:24:17.00 &  II & binary ($i\sim 85^{\circ}$ $^{\mathrm{h}}$)\\
UY Aur      & 04:51:47.31, +30:47:13.9 & II & $i\sim 42^{\circ}$ \\
\hline                                   %inserts single line
\end{tabular}
\begin{list}{}{}
\item[$^{\mathrm{a}}$] Coordinate of the center of the 1' aperture.
\item[$^{\mathrm{b}}$] Inclination of the disk estimated by Tobin et al. (\cite{tobin08}).
\item[$^{\mathrm{c}}$] Judged from the similarity of its N3 image and spectrum to those of L1527 and IRAS04302. 
\item[$^{\mathrm{d}}$] Wolf, Padgett, \& Stapelfeldt (\cite{wolf03})
\item[$^{\mathrm{e}}$] Hodapp et al. (\cite{hodapp04})
\item[$^{\mathrm{f}}$] Grosso et al. (\cite{grosso03})
\item[$^{\mathrm{g}}$] Inclination of HV Tau C (Monin \& Bouvier \cite{monin2000})
\item[$^{\mathrm{h}}$] Inclination of HK Tau B (Stapelfeldt et al. \cite{stapelfeldt98})
\end{list}

\end{table*}

\begin{figure*}
\centering
\includegraphics[width=12cm]{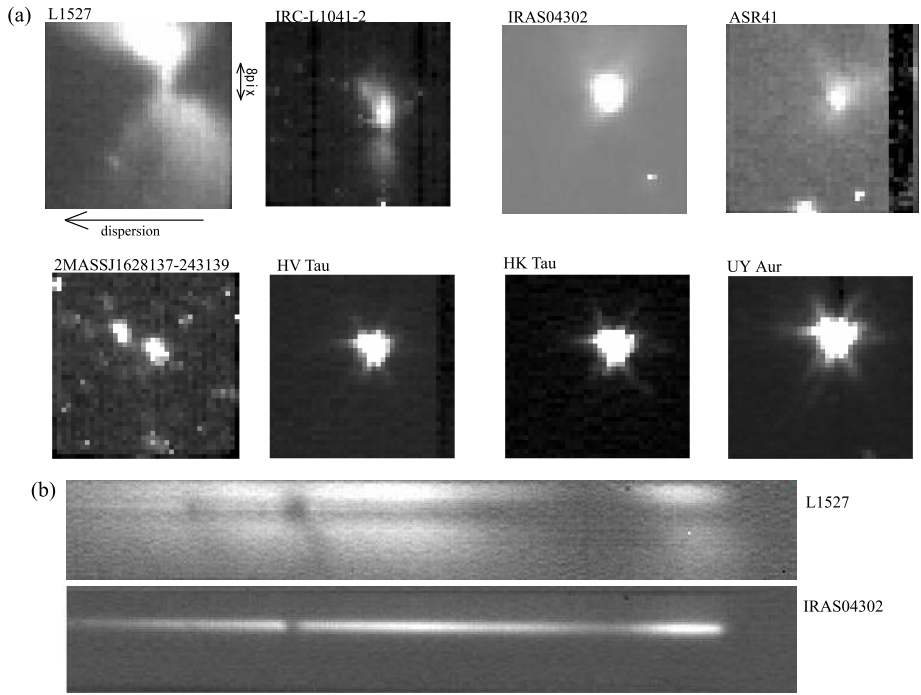}
   \caption{(a) Images of our targets taken by AKARI with the N3 filter ($2.7-3.8 \mu$m).
            The stellar light is dispersed from right (short wavelength) to left (long wavelength).
            The FOV is $1'\times 1'.$
            The coordinate at the center of the FOV is listed in Table \ref{targets}. Images of
           IRC-L1041-2 and 2MASS J1628137-243139 are damaged by cosmic rays, while
           images of HV Tau, HK tau, ASR41, and UY Aur are affected by saturation and column pull down.
           (b) Spectral images of 1527 and IRAS04302.}
      \label{N3}
\end{figure*}

\section{Data reduction and analysis}
\subsection{Reduction of the spectral image}
The reduction pipeline provided by the AKARI project team, {\it IRC$\_$SPECRED} 
(Ohyama et al \cite{ohyama07}), is unsuitable for our targets, because it
automatically defines the area surrounding selected sources
as ``sky'', an area that is often contaminated by nebulosity in our observations.

To derive spectra from the data, we developed our own procedure, referring to the reduction 
method by Sakon et al. (\cite{sakon}). Firstly, we made a cosmic-ray-free dark for each target by taking the median
of three dark frames; two dark frames were taken during the observation of each target, and one additional
dark frame was taken from the preceding or subsequent observation in the AKARI archive. This procedure was
necessary, because some frames are heavily damaged by cosmic rays. The dark was subtracted
from the N3 image and spectral images.

The eight spectral images were slightly shifted in relation to each other by the attitude drift of the satellite. 
The task {\it IRC\_SPECRED}
measures the shift of each image relative to the fourth image. Referring to this measurement,
we selected images with a relative shift of within 0.5 pixels, in both spatial and spectral
directions. Images with larger shifts were discarded.
Cosmic rays were removed by subtracting the median of the possible combinations of
three frames from the selected frames. We then stack and average these cosmic-ray-free spectral images.
%We also applied stacking procedure developed by \cite{noble09}; each spectral image is shifted
%in spatial direction via interpolation to coincide with the 4th image. We confirmed that
%the spectra derived by these two stacking methods are not significantly different.

The local-sky region was selected by carefully avoiding nebulosity, field stars, and bad pixels.
In the case of L1527, there is no 'sky' region, because the nebulosity extends all over
the 1' $\times$ 1' aperture (Fig \ref{N3}). Since we performed slitless spectroscopy,
the spectral image of the nebulosity is a convolution of the nebula spectrum with the brightness
distribution, and should not be considered as 'sky'. We selected from the AKARI
archive a 1' $\times$ 1' spectral image that can be used as sky; i.e. a dark
region observed at a similar period and coordinate to those of L1527.

After the sky subtraction, we defined the spectrum region of each star in the spectral image,
referring to the coordinates of the objects and the N3 image. The signal was integrated over
eight pixels (12 arcsec) in the spatial direction. As an alternative approach, Noble et al. (in prep) fit the
signal at each wavelength with the PSF and integrated the signal. We confirmed that the result
of the eight-pixel integration agrees well with this more sophisticated approach. 
We note that the spectrum is not perfectly aligned
with the rows of the pixels. We adopted the formulation as the tilt of the spectrum
\begin{equation}
Y=-0.00746929 (X-X_0) +Y_0,
\end{equation}
where X and Y are the pixel number in the dispersion direction and spatial direction respectively, and
$Y_0$ is the center of the spectrum at an arbitrary position, $X_0$, in the dispersion direction.

In Fig \ref{N3}, nebulosity is clearly seen towards L1527, IRC-L1041-2, IRAS04302, and ASR 41.  
For these objects, we set $Y_0$ in Eq. (1) as the brightest position in the spatial
direction at a wavelength ($X_0$) in the spectrum image, and integrated the spectrum over eight pixels
in the spatial direction at each wavelength (Figure \ref{N3}).
One exception is L1527; since the central point-like source is apparent and fainter than the nebulosity,
we set $Y_0$ as the central (faintest) position in the spatial direction, and integrated the signal
over four pixels in the spatial direction (Figure \ref{N3}).
%Aperture correction factor for the 8 pixel integration is $0.98958 - 0.02028 \lambda$.

The integrated signal was divided by the aperture correction factor and response function before being converted
into a spectrum. The response function was determined by the observation of a standard star (KF09T1).
Wavelength calibration was done by the AKARI project team using the slit spectroscopy observation
of standard stars. In our observations, we determined the wavelength of the spectrum by measuring
the shift between the slit and the direct light position of the target in the N3 image.

According to the IRC team, the pixel-wavelength relation is a linear function. This linear correlation is
confirmed in the wavelength range 2.5-4.5 $\mu$m, although the calibration is less certain at longer
wavelengths. The CO absorption peak ($\sim 4.67$ $\mu$m) is shifted to shorter wavelengths by 1-2 pixels
in our data. Since similar shifts in this wavelength are reported in other IRC data, we shifted the data at $\ge
4.5$ $\mu$m by 1-2 pixels towards longer wavelengths.

The error in flux was evaluated as follows. Firstly, we estimated the noise level in the 'sky' region and converted it
into an error in flux using the response function. Secondly, we estimated the flux errors caused by the
wavelength assignment; we calculated fluxes by assuming that the shift between the slit and the direct light position
of the target is 0.5 pixels larger/smaller. We then adopted the maximum of these errors. 

\subsection{Absorption features}

To evaluate the ice absorption profile quantitatively, we normalized the spectrum with
the continuum, which is assumed to be a second order polynomial, and was determined by referring to the
three wavelength regions $\sim 2.6$ $\mu$m, $3.4-3.8$ $\mu$m, and $4.2-4.9$ $\mu$m. The choice of
 wavelengths differed slightly among objects to ensure a good fit to all spectra.

Template absorption profiles were obtained from ``data base 2007''
at Raymond \& Beverly Sackler Laboratory for Astrophysics\footnote{
http://www.strw.leidenuniv.nl/$^{\sim}$lab/}.
The database summarizes the absorption profiles of ice species measured in laboratories.
Gerakines et al. (\cite{gerakines1995}; \cite{gerakines1996}) measured the profiles and determined
band strengths of various astrophysical ice analogs. 
From the database of Gerakines et al. (\cite{gerakines1995}; \cite{gerakines1996}), 
we adopted profiles of pure ices at 10 K unless otherwise stated.
In reality, interstellar ice is a mixture of various components such as
H$_2$O, CO, and CO$_2$, and the peak position and shape of the absorption features change with
the abundance ratio of the components. However, our spectral resolution is insufficiently high to perform
such a detailed analysis of the ice mixture.

The profiles of the 4.27 $\mu$m CO$_2$ band
and the 4.67 $\mu$m CO band depend strongly on grain shape (Ehrenfreund et al \cite{ehrenfreund1997}).
We adopted the continuously distributed ellipsoids (CDE) grain model for the CO band feature
(Ehrenfreund et al \cite{ehrenfreund1997}, Boogert, Blake \& Tielens \cite{boogert2002},
Pontoppidan et al. \cite{pontoppidan2003}).
For the CO$_2$ band, on the other hand, we adopted the spectrum of Gerakines et al. (\cite{gerakines1995}),
which is uncorrected for the grain shape, since it gives a good fit to our observational data.
The pure CO$_2$ feature with grain-shape (CDE) corrections agrees with neither our data nor the CO$_2$ ice feature
observed by ISO towards high-mass YSOs, which indicates that CO$_2$ ice is not pure in interstellar medium.
The ISO observations of the 4.27 $\mu$m feature indicated that the CO$_2$ ice is primarily in a polar (H$_2$O-rich)
component of the ices (Whittet et al. \cite{whittet98}). However, to interpret
our observation, we also need to introduce an apolar component to reproduce the blue wing at $\sim 4.2$ $\mu$m.
Since our spectral resolving power is low, we simply use the spectrum of Gerakines et al. (\cite{gerakines1995})
as a template, rather than combining multiple components of ice mixture with grain shape corrections.
We also adopted the 3.05 $\mu$m H$_2$O band of Gerakines et al (\cite{gerakines1995}), since it is little affected
by the grain shape.

The spectral resolution of IRC, $\sim 100$, is not high enough to fully resolve
narrow bands such as CO and CO$_2$.
We fitted these narrow bands using the following method: we considered the absorbance of ice from
the database and convert it into a normalized flux spectrum, assuming the optical depth of ice along the line of sight.
The spectrum was then convolved with the instrumental line profile and re-binned to the resolving power of
AKARI, to be compared with the observed spectrum. We vary the optical depth of ice until the best-fit is
found by least squares fitting. Broad absorption features such as the H$_2$O band at 3.05 $\mu$m were
not significantly altered by the convolution.

In N3 images ($2.7-3.8 \mu$m), class 0-I objects (L1527, IRC-L1041-2, and IRAS04302) are more extended than a point source.
Among the three objects, L1527 is the most extended, and IRAS04302 is the least extended. Since the 
IRC spectrograph is slitless, the spectral resolving power depends sensitively on the spatial extent of the source. Assuming
that the projected column density of ice in front of the extended sources is constant, the observed band profile
is a convolution
of the ice absorptions with the spatial distribution of the light source in the dispersion direction.
For L1527, we integrated the central four pixels of the N3 image in the spatial direction around the source position
to derive the spatial distribution of the light source in the dispersion direction (N3 profile, hereafter),
with which the laboratory ice spectrum was convolved. 
This convolution better reproduces the observed features of CO and CO$_2$ than the convolution
with the instrumental line profile. For IRC-L1041-2 and IRAS04302,
however, we convolved the ice features with the instrumental line profile, which reproduces the observed spectrum better
than the N3 profile. Since the instrumental line profile at 4 $\mu$m (N4 band) is
narrower than that of the N3 band, the N3 profile might be too broad
to be applied to CO and CO$_2$ features. Unfortunately, imaging observation was not performed in the N4 band.

%One exception is the 3.05 $\mu$m H$_2$O band; it is so broad that we do not need
%to convolve the spectrum with the instrumental line profile. It should be noted that we use the normalized flux
%instead of optical depth for the least square fitting, because our CO and CO$_2$ features are broadened by the PSF,
%and because the 3.05 $\mu$m water band of some targets is saturated. 

Once we determined the best-fit absorption feature, we used the original, pre-convolution profile to calculate the
the column density $N$ of ice
\begin{equation}
N=\int \frac{\tau d\nu}{A},
\end{equation}
where $A$ is the band strength (e.g. Whittet et al. \cite{whittet93}). We adopted the band strength listed in Table 2 
of Gibb et al. (\cite{gibb04}), unless otherwise stated (Table \ref{column}).
The CO ice column density of the CDE grain model was calculated to be
\begin{equation}
N=6.03 \tau_{\rm max} A^{-1},
\end{equation}
where $A=1.1\times 10^{-17}$ cm molec$^{-1}$ and $\tau_{\rm max}$ is the optical depth at the
wavelength of the absorption peak
(Pontoppidan et al. \cite{pontoppidan2003})

Since the spectral resolution is low, and since some lines are saturated, it was difficult to estimate the
errors in the ice column densities accurately. In the present paper, the column density error was evaluated by
simply calculating the error in the absorption area. For CO$_2$, we calculated the absorption area $\int \tau d\nu$
at $4.2-4.3$ $\mu$m and its error caused by the flux error. Assuming that the absorption area had an error of 11 \%, and
that the
CO$_2$ column density was estimated to be $10.0 \times 10^{17}$ cm$^{-2}$, for example, the column density error
was calculated as $1.1\times 10^{17}$ cm$^{-2}$.
For H$_2$O, we estimated the error by calculating the absorption area between $2.8-3.2$ $\mu$m.
If the red wing was much broader than the laboratory spectrum, we estimated the error from the absorption area
at $2.8-3.0$ $\mu$m. For L1527, towards which the H$_2$O band is saturated, we used the wavelength range
$\sim 2.8-2.9$ $\mu$m. The absorption at $\sim 4.7$ $\mu$m
is a combination of the absorptions of CO, XCN, and CO gas (\S 4.5).
Here, the error is evaluated by calculating the absorption area at $4.65-4.7$ $\mu$m for CO
and  $4.57-4.65$ $\mu$m for XCN.
Uncertainties in ice column densities originating from the choice of the convolution function
(instrumental line profile or N3 profile) are discussed in \S 4.

The derivation of ice column densities described above implicitly assumes a simple geometry: a point light source behind 
a uniform absorbing material. Edge-on YSOs have more complicated structures. Stellar light scattered by the outflow cavity
can be an extended light source (e.g. Tobin et al. \cite{tobin08}), even if it looks like a point source with a PSF of
a few arc seconds, and the column density of the absorbing material (envelope and/or disk) also varies spatially.
More accurate estimates of the ice column densities requires the radiative transfer calculation of two-dimensional (2D) or
three-dimensional (3D)
models of envelope and disk structure (Pontoppidan et al. \cite{crbr05}). We compare the observed spectra with existing
model calculations and discuss the location of ice in \S 5.

\section{Results}
\subsection{Spectrum}
Figure \ref{spec} shows the derived spectra of our target YSOs. 
%Red crosses in the panels of L1527, IRAS04302, ASR41 and 2MASSJ1628137-243139
%depict the photometry values measured with
%%2MASS (1.235 $\mu$m, 1.662 $\mu$m, and 2.159 $\mu$m) and 
%the {\it IRAC} (Channel I and II) on board {\it SST}. Although we plot the {\it IRAC} data at 3.6 $\mu$m
%and 4.5 $\mu$m, the {\it IRAC} actually covers  $3.2$ $\mu$m$-3.9$ $\mu$m (Channel I) and
%$4.0$ $\mu$m$-5.0$ $\mu$m (Channel II). 
%The pixel scale of {\it AKARI} is 1.46" at NIR, while that of
%{\it IRAC}  Channel I and II are 1.221" and 1.213", respectively. Since we integrate the {\it AKARI}
%spectral image over 8 pixels ($11.68"$) in spatial direction, we integrated over the radius of 5 pix
%($\sim 6"$) to derive the photometry values from {\it IRAC} image.
%The fluxes of our spectra show reasonable agreement with the photometry data.
The spectrum of L1041-2 resembles that of L1527, which suggests a similarity in the structure
and evolutionary stage of these objects.  The spectrum of IRAS04302, on the other hand, is less
reddened than L1527. It might be in a more evolved stage than L1527 and L1041-2.

2MASS J1628137-243139 is very faint and is almost at the faintest limit for which we can derive a spectrum.
The spectrum of ASR 41 shows a sharp decline at $\sim 4.5$ $\mu$m and is noisy at
longer wavelengths because of the column pulldown caused by a bright star in the 10' $\times$ 10' aperture.

HV Tau is a triplet system:
HV Tau A and HV Tau B are a binary system with a separation of 0.0728" (Simon et al. \cite{simon98}), and our target edge-on
disk, HV Tau C, is separated by $\sim 4"$ from the binary (Woitas \& Leinert \cite{woitas98}).
Since the FWHM of the AKARI PSF is $\sim 4.3"$ in the NIR and the PSF is not circular,
it is difficult to disentangle the HV Tau C spectrum from that of the system. Hence, we integrated the spectral image
over an eight-pixel width. The spectrum is dominated by HV Tau A and B at $2.5-4.5$ $\mu$m,
as they are brighter than HV Tau C by 4.22 mag in the K band and $>1.71$ magnitude in the L band.
At $\>4.5$ $\mu$m, on the other hand, the contribution of HV Tau C may not be negligible, because
HV Tau C is as bright as the HV Tau AB system in the N band (Woitas \& Leinert \cite{woitas98}).
HK Tau is also a binary system. The edge-on object, HK Tau B, is a faint binary.
Unfortunately, it is impossible to disentangle the spectrum of HK Tau B from the primary,
because the two stars almost align in the dispersion direction. It is disappointing that we were unable
to discern the spectra of HV Tau C and HK Tau B.
The absorption features are, however, still apparent in our spectra. We compare these absorptions
with Subaru observations towards the HV Tau C and HK Tau B by Terada et al. (\cite{terada07}) in the following discussion.

The dashed lines in Fig \ref{spec} depict the continuum.
While the continuum varies between objects, reflecting their structure and evolutionary stage, we can clearly see absorption bands of H$_2$O (3.05 $\mu$m),
CO$_2$ (4.27 $\mu$m), and CO (4.67 $\mu$m) in most of the spectra. The assignment and fitting of
each ice band are performed in the following subsections.

\begin{figure}
\centering
\includegraphics[width=9cm]{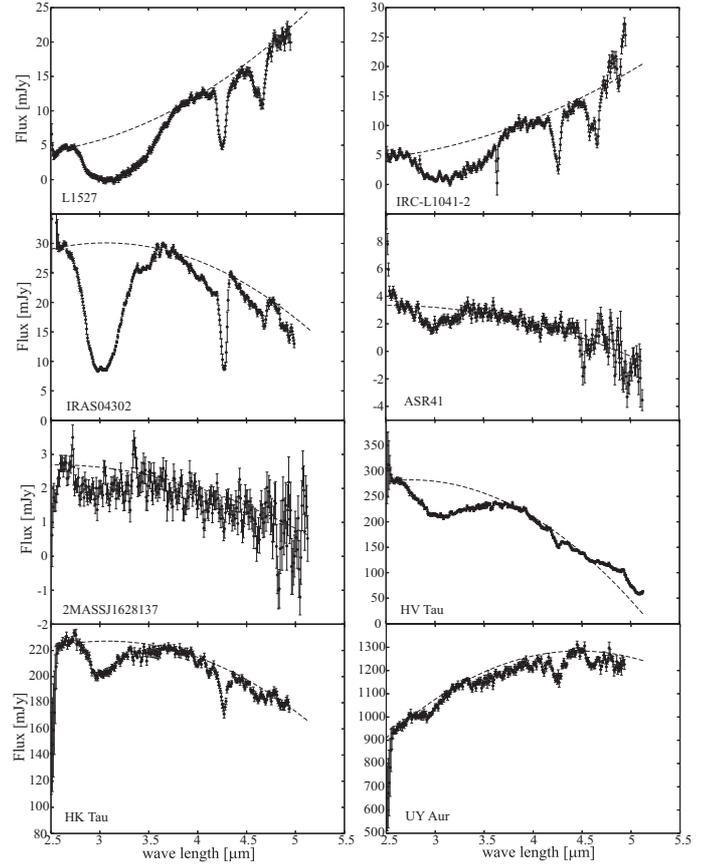}
   \caption{The derived spectra of target YSOs in the NIR (closed circles with error bars). The continuum (dashed line)
                is determined by the second-order polynomial fitting of the spectrum. %Red crosses depict
                %the flux measured with Channel I (3.6 $\mu$m) and Channel II (4.5 $\mu$m) of IRAC. 
                 }
      \label{spec}
\end{figure}

\subsection{H$_2$O ice}
Figure \ref{norm_H2O} shows the spectra normalized by the continuum. The vertical axis is the optical
depth of the absorption.
Towards L1527 and IRC-L1041-2, the 3.05 $\mu$m water band is clearly saturated; the flux at the
absorption peak is almost zero (Figure \ref{spec}). We fitted the spectra in the range $2.9-3.1$ $\mu$m with 
the laboratory spectrum of pure H$_2$O ice at 10K 
(Gerakines et al. \cite{gerakines1995})(red lines in Figure \ref{norm_H2O}).
We also fitted the observed feature with the pure H$_2$O ice feature at 160 K. The agreement is much
worse than for the ice at 10 K, which suggests that the bulk of the observed H$_2$O ice is not annealed.

Interestingly, the profiles fitted at $2.9-3.1$ $\mu$m (solid lines in Figure \ref{norm_H2O}) 
reproduce the shallow hollow at 4.5 $\mu$m, which is the O-H combination mode
(Boogert et al. \cite{boogert2000}). While the 3.05 $\mu$m
band is saturated, the 4.5 $\mu$m band suggests that the H$_2$O column density cannot be significantly
larger than we estimated.

Table \ref{column} summarizes the column density of H$_2$O ice derived by fitting the 3.05 $\mu$m band.
The class 0-I objects L1527, IRAS04302, and IRC-L1041-2 clearly have much higher water column densities
than the other targets, probably because of the larger mass of their circumstellar material and geometrical effects (\S 5).

The observed profiles show a red wing ($> 3.1$ $\mu$m), which
is often observed in molecular clouds. The shape and
depth of the red wing relative to the peak (3.05 $\mu$m) vary significantly among objects;
it is especially deep towards L1527 and L1041-2.
The wing is generally attributed to the scattering by large grains with ice mantles
(e.g. L\'{e}ger et al. \cite{leger1983}). The size of the grains responsible for the red wing
is $\lambda/2\pi\sim 0.5$ $\mu$m.

ASR 41 has a double-peaked profile at the absorption peak $\sim 3$ $\mu$m. We reduced each
frame of the spectral image to confirm that the double-peaked profile is not caused by noise in
a specific frame. The double-peaked profile remained, even if we changed the combination of
frames in the stacking. We fitted the profile with (i) the H$_2$O band alone (red line
in Figure \ref{norm_H2O}) and (ii) the H$_2$O band plus an NH$_3$ band (2.98 $\mu$m) (blue line in
Figure \ref{norm_H2O}). Although the blue line seems to fit the observed profile slightly
better, it results in an unreasonably high column density of NH$_3$
($2\times 10^{18}$ cm$^{-2}$) compared with the H$_2$O column density
($7.8 \times 10^{17}$ cm$^{-2}$), and the bump at 3.1 $\mu$m is not fitted well.
While the double-peaked profile seems to be robust,
it could still be noise, since the flux at the absorption peak is only $1-2$ mJy.
We therefore prefer to fit the 3 $\mu$m absorption profile with the H$_2$O band alone.

In the polynomial fitting of the HV Tau continuum, we ignored the spectrum at $>4.5$ $\mu$m,
since the gradual rise of the spectrum at longer wavelengths may be caused by HV Tau C, while
shorter wavelengths are dominated by HV Tau AB.
The absorption band at 3 $\mu$m is much
broader than the laboratory profile. While the red wing may be caused by scattering,
the blue wing may be caused by the combination of the spectra of HV Tau A
and B. The binary is close enough ($\sim 0.07"$) to be considered as a point source in our
observation, but the objects have similar brightnesses and different colors (Simon et al \cite{simon98}).
We tentatively fitted the absorption peak with the laboratory data to derive the water ice column
density.

Terada et al (\cite{terada07}) detected deep H$_2$O absorption towards the edge-on disks of HV Tau C
and HK Tau; they concluded that these absorptions originate in the disk ice. 
If there were no foreground clouds with ice, our spectra should be a combination of the edge-on disk with ice features
and bright primaries without absorption.
The observed 3 $\mu$m absorption of HV Tau is, however, too deep to be accounted for by the disk
ice alone; while the flux of HV tau C is 7.75 mJy at K band and 7.14 mJy at L' band, the 3 $\mu$m absorption is
as deep as $> 50$ mJy. There must be a contribution of ice absorption in foreground clouds.
The H$_2$O column density in Table \ref{column} should be considered as upper limits to the ice in
foreground clouds, since we do not distinguish between the disk ice and foreground ice.
However, the 3 $\mu$m absorption of HK tau might be due to the disk ice of HK tau B;
the flux of HK tau B is 15.9 mJy at 2.2 $\mu$m
and 9.85 mJy at 3.8 $\mu$m, while the absorption is $\lesssim 20$ mJy. 
If we were to assume that the flux of the primary is 200 mJy, subtract it from the spectrum, and fit the 3 $\mu$m absorption
by the H$_2$O ice feature, the ice column density would be $3 \times 10^{18}$ cm$^{-2}$, which is consistent with the
value derived by Terada et al. (\cite{terada07}) ($3.31 \times 10^{18}$ cm$^{-2}$).

%\begin{table*}
%\caption{Ice column density towards objects}
%\label{column}
%\centering
%\begin{tabular}{l c c c c c}
%\hline\hline 
%target & N(H$_2$O) & N(HDO) & N($^{12}$CO$_2$)  & N(CO)  & N(XCN) \\
%         & $10^{17}$ [cm$^{-2}$] & $10^{17}$ [cm$^{-2}$] &  $10^{17}$ [cm$^{-2}$] &  $10^{17}$ [cm$^{-2}$] 
%        & $10^{17}$ [cm$^{-2}$] \\
%\hline                        % inserts single horizontal line
%%                                    H2O  HDO CO2    CO     COgas   XCN
%L1527(04368+2557)         &49 & 0.27&   7.2 &  7.9    &  1.8  \\
%IRC-L1041-2                  & 39 & 0.79&   8.1 & 18.0 &  2.5  \\
%IRAS04302+2247             & 24 & 1.3 &  6.8 & 3.1  &   0.53   \\
%ASR 41                           &7.8 & - &   -  & -      &   -   \\
%2MASSJ1628137-243139 & 6.7&  - &  -  & -      &   -  \\
%HV Tau                           &5.4& 0.26 & 0.72&  -    & -   \\
%HK Tau                           & 2.1 & -&   0.9&  -      & -   \\
%UY Aur                           & 0.61&-&0.54 &      & - \\
%\hline                           %inserts single line
%\end{tabular}
%\end{table*}

\begin{table*}
\caption{Absorption peak, band strength, and column densities of ices towards objects}
\label{column}
\centering
\begin{tabular}{l c c c c c}
\hline\hline 
                          & H$_2$O & HDO & $^{12}$CO$_2$  & CO  & XCN \\
\hline
Absorption peak [$\mu$m]&  3.05 & 4.07 & 4.27 & 4.67 & 4.62 \\
Band strength $10^{-17}$ cm molecule$^{-1}$& 20           & 4.3 & 7.6 & see text &  5.0 \\
%$10^{-17}$ cm molecule$^{-1}$ & & & & & \\
\hline\hline
Column density   & $10^{17}$ [cm$^{-2}$] & $10^{17}$ [cm$^{-2}$] &  $10^{17}$ [cm$^{-2}$] &  $10^{17}$ [cm$^{-2}$] 
        & $10^{17}$ [cm$^{-2}$] \\
\hline                        % inserts single horizontal line
%                                    H2O                      HDO                  CO2                            CO                     XCN
L1527(04368+2557)         &47$\pm 13$  & 0.96 ($2.0\%$)$^{\mathrm{a}}$&  10$\pm 1.1$ (21\%) &  18$\pm 2.6$ (38 \%) & 1.5 $\pm 0.22$ (3.2 \%) \\
IRC-L1041-2                  & 39$\pm 6.6$ &  3.7 ($9.5 \%$) & 9.4$\pm 1.6$ (24 \%) & 18$\pm 4.0$ (46\%) &  2.5$\pm 0.48$ (6.4 \%) \\
IRAS04302+2247             & 24$\pm 0.82$ & 5.3 ($22 \%$) &  6.8$\pm 0.64$ (28 \%)& 3.1$\pm 0.55$ (13 \%)  &   0.53$\pm 0.064$ (2.2 \%)  \\
ASR 41                           &7.8$\pm 2.2$ & - &   -  & -      &   -   \\
2MASSJ1628137-243139 & 6.7$\pm 2.6$&  - &  -  & -      &   -  \\
HV Tau                           &5.4$\pm 0.23$& 1.0 ($19 \%$) & 0.72 $\pm 0.11$ (13 \%)&  -    & -   \\
HK Tau                           & 2.1$\pm 0.23$ $^{\mathrm{b}}$ & -&   0.9$\pm 0.11$ (43 \%) &  -      & -   \\
UY Aur                           & 0.61$\pm 0.20$&- & 0.54$\pm 0.086$ (89 \%) &      & - \\
\hline                           %inserts single line
\end{tabular}
\begin{list}{}{}
\item[$^{\mathrm{a}}$]  Abundance relative to H$_2$O ice.
\item[$^{\mathrm{b}}$] If the absorption originates in HK Tau B, ice column density is $3\times 10^{18}$ cm$^{-2}$ (see text).
\end{list}
\end{table*}

\begin{figure}
\centering
\includegraphics[width=9cm]{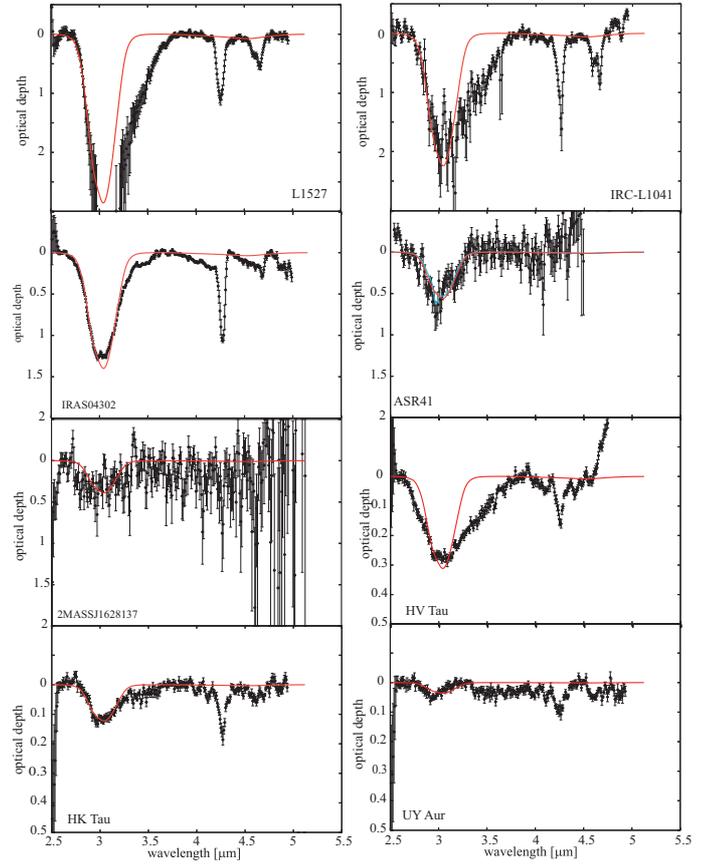}
   \caption{Closed circles with error bars depict the YSO spectrum normalized by the continuum.
                The absorption feature at $3.05 \mu$m is fitted by the laboratory spectrum of pure H$_2$O ice
                at 10 K (red line) (see text).}
      \label{norm_H2O}
\end{figure}

\subsection{CH band at 3.5 $\mu$m}

While the long wavelength wing of H$_2$O absorption is generally attributed to scattering
by large grains with ice mantles, it might also represent the stretching mode of C-H bond. Among various possible
carriers of the C-H bond, CH$_3$OH ice has long been observed via the substructures at
3.47 $\mu$m and 3.53 $\mu$m (e.g. Grim et al. \cite{grim1991}; Brooke, Sellgren \& Smith \cite{brooke1996}).
While the earlier works are mostly restricted to high-mass and intermediate YSOs,
Pontoppidan et al (\cite{ponto_ch3oh}) identified the
3.5 $\mu$m feature towards three low-mass YSOs (out of 40 sources) as CH$_3$OH ice. 
%Since the detection of CH$_3$OH towards low-mass YSO is very limited,
Boogert et al. (\cite{boogert08}) detected the C-O stretching mode of CH$_3$OH at 9.7 $\mu$m
towards 12 low-mass YSOs (out of 41 sources) using Spitzer.
Boogert et al. (\cite{boogert11}) detected the 3.53 $\mu$m band of CH$_3$OH towards
several field stars behind isolated dense cores; the CH$_3$OH/H$_2$O ratio ranges from 5 \% to 20 \%.
The ratio of CH$_3$OH to H$_2$O column densities varies considerably among objects.

Another possible carrier of the C-H bond is CH$_4$, whose absorption bands at 3.32 $\mu$m and 7.67 $\mu$m have been 
detected towards high-mass YSOs (e.g. Lacy et al. \cite{lacy91}; Boogert et al. \cite{boogert04}).  {\"O}berg et al.
(\cite{oberg_ch4}) detected a 7.7 $\mu$m absorption feature, which is attributed to CH$_4$, towards 25 sources out of 52
low-mass YSOs.
Since CH$_3$OH and CH$_4$ are considered to be the precursors of large organic molecules in hot corinos and WCCC,
respectively, it is important to constrain their abundances using the spectrum around 3.5 $\mu$m.

%L1527, a WCCC object, is an important object towards which the 3.5 $\mu$m feature should be
%investigated. While CH$_4$ is the precursor of carbon chains (e.g. Aikawa et al. \cite{aikawa08}
%and references therein), methanol is considered to be a precursor
%of large organic species. Since the sublimation temperature of CH$_3$OH ($\sim 100$ K) is
%higher than that of CH$_4$ ($\sim 25$ K),
%it is important to constrain the solid CH$_3$OH abundance towards WCCC objects, in order to clarify
%whether the non-detection of large organic species is due to low temperatures ($< 100$ K) 
%or peculiar molecular abundances with little CH$_3$OH.

Figure \ref{CH_3um} shows the spectra of L1527, IRAS04302, and HK Tau around 3.5 $\mu$m.
While the spectrum of L1527 is smooth, those of IRAS04302 and HK Tau contain absorption features. These features,
however, do not match either CH$_3$OH or CH$_4$.
A potential complication to the identification of, in particular, the CH$_4$ 3.32 $\mu$m band and to a lesser
degree the CH$_3$OH 3.53 $\mu$m band, is the possible presence of the 3.3 $\mu$m PAH emission feature and the
HI Pf$\delta$ line at 3.297 $\mu$m. We can indeed see an emission feature at 3.3 $\mu$m in HK Tau that could
be ascribed to the PAH band. In the following, we derive upper limits to the column densities of CH$_3$OH and
CH$_4$, keeping in mind that the low spectral resolving power of our spectra may blend 
adjacent emission and absorption bands. 
%\bf Another complication is that the emission lines of HI, H$_2$ and PAH are detected towards some
%low-mass YSOs (e.g. Geers et al. \cite{geers07}) at these wavelengths.
%For example, Pf $\delta$ and PAH emissions are at $\sim 3.3$ $\mu$m,
%at which we see a bumpy feature in HK Tau. In the following, we derive the upper limits of
%CH$_3$OH and CH$_4$ by comparing the observed features with the laboratory ice spectra, although the observed
%feature could be contaminated by the emission lines, which is difficult to identify with our low spectral resolution.

The thick solid lines in Figure \ref{CH_3um} depict the upper limits on the CH$_4$ absorption depth.
We convolved the pure CH$_4$
ice feature at 10 K (Gerakines et al. \cite{gerakines1995}) with the instrumental
line profile (IRAS04302 and HK Tau) or the N3 profile (for L1527).
The upper limits to the CH$_4$ column density are $1.6 \times 10^{19}$ cm$^{-2}$ (larger than the H$_2$O column density),
$3.1 \times 10^{17}$ cm$^{-2}$ (13 \% relative to H$_2$O), and $1.6 \times 10^{17}$ cm$^{-2}$ (76\%) towards L1527, IRAS04302,
and HK Tau, respectively. Owing to the low flux, the upper limit towards L1527 is especially high. 

The dashed lines show the upper limits to the CH$_3$OH absorption depths, using a laboratory spectrum for pure CH$_3$OH at 10K
(Gerakines et al. \cite{gerakines1995}), convolved with the instrumental line profile. While the 3.53 $\mu$m feature is common
among pure CH$_3$OH ice and mixed ice (H$_2$O+CH$_3$OH), the shape and depth of the 3.4 $\mu$m feature
depend on the composition of the mixture (e.g.  Pontoppidan et al. \cite{ponto_ch3oh}). Hence, we derived
upper limits to the CH$_3$OH column density of the 3.53 $\mu$m feature. The upper limits to the CH$_3$OH column densities are
$1.2 \times 10^{18}$ cm$^{-2}$ (26\% relative to H$_2$O), $1.0 \times 10^{18}$ cm$^{-2}$ (42\%), and
$3.1 \times 10^{17}$ cm$^{-2}$ (larger than the H$_2$O column) towards L1527, IRAS04302, and HK Tau, respectively.

%CH4
%1.57E+17 HK Tau
%1.57E+19 L1527
%3.14E+17 IRAS04302

%CH3OH
%3.100E+17 HK Tau
%1.1625E+18 L1527
%1.0075E+18 IRAS04302

%The laboratory
%pectra of CH$_4$, CH$_3$OH and CH$_3$CH$_2$OH are shown in Figure \ref{CH_3um} (c)
%(Gerakines et al. \cite{gerakines1995}). IRAS04302 shows absorption at $3.35 - 3.5$ $\mu$m,
%where CH$_3$OH and CH$_3$CH$_2$OH absorb. The observed feature, however, does not
%match with either of the laboratory spectra. L1527, on the other hand, has a very smooth spectrum
%without any narrow features attributable to CH$_3$OH or CH$_4$. Since the flux at $\sim 3.5$ $\mu$m is low,
%the upper limit on the ice column densities is not very strict: $N$(CH$_4$)$\le 9 \times 10^{17}$ cm$^{-2}$
%and $N$(CH$_3$OH) $\le 3.8 \times 10^{17}$ cm$^{-2}$.

%CH3CH2OH+ H2O: 3.355, 3.44, 3.477
\begin{figure}
\centering
\includegraphics[width=7cm]{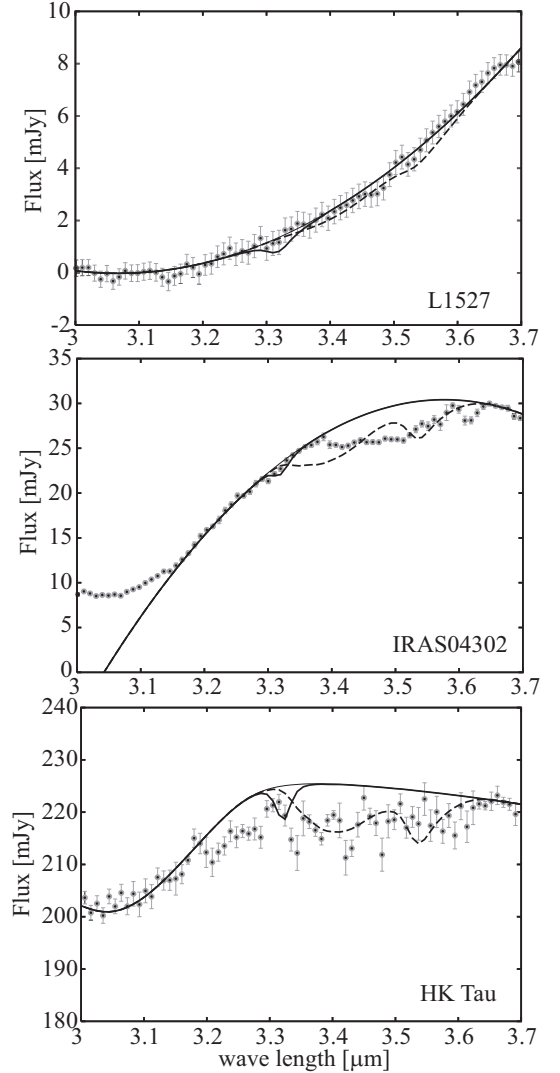}
   \caption{Three-micron spectra of L1527, IRAS04302, and HK Tau (gray dots with error bars). The thin solid lines in
   L1527 and IRAS04302 are the local (3.1-3.7 $\mu$m) continua fitted by second order polynomials, while the line
   in HK Tau is the best-fit absorption of H$_2$O (Figure \ref{norm_H2O}). Solid and dashed lines
    depict the upper limits to the pure CH$_4$ ice and pure CH$_3$OH ice absorption depth, respectively.}
     \label{CH_3um}
\end{figure}

\subsection{HDO ice}
In some objects, such as IRAS04302, we can see absorption around 4.1 $\mu$m, where the
OD stretching mode of HDO is observed in the laboratory (Dartois et al. \cite{dartois03}).
Since the detection of HDO ice is very rare (Teixeira et al. \cite{teixeira99}) and has not been confidently
confirmed, we checked the response function carefully and confirmed that the $\sim 4.1$ $\mu$m
feature is not caused by an artifact in the response function. 
Figure \ref{HDO} shows the spectra of this wavelength region towards L1527, IRC-L1041-2, IRAS04302,
and HV Tau, which are fitted with a model of the amorphous HDO feature at 10 K: a Gaussian profile
peaking at 4.07 $\mu$m with a full-width half maximum (FWHM) of 0.2 $\mu$m (solid lines). The spectrum of L1527 is not
fitted well with this Gaussian; the absorption has a peak at longer wavelength ($\sim 4.13$ $\mu$m)
and a narrower band width, which resembles an annealed, rather than an amorphous, HDO feature
(Dartois et al. \cite{dartois03}). We fitted the L1527 spectrum with a Gaussian peaking at 4.13 $\mu$m
and FWHM of 0.1 $\mu$m (dashed line in Figure \ref{HDO}).

Although IRAS04302 has the deepest and smoothest absorption around 4.1 $\mu$m, this result should be taken
with caution. The spectrum also shows a broad absorption at $4.5-4.6$ $\mu$m, which cannot be
fitted well by the absorptions of CO and XCN (see \S 4.6). These two absorptions (4.1 $\mu$m and $4.5-4.6$ $\mu$m) could
be related; there could be an alternative explanation, rather than HDO, CO, and XCN. The features themselves are, however,
robust. We have two independent data sets of IRAS04302, and the two broad absorptions appear in both
data sets.

We integrated the fitted spectra to derive the column density of HDO (Table \ref{column}).
The band strength was assumed to be $4.3 \times 10^{-17}$ cm molecule$^{-1}$ (Dartois et al. \cite{dartois03}).
Considering the small bumps and hollows that deviate from the HDO feature and the above discussion of
IRAS04302 spectrum, the HDO column densities should be interpreted with caution. However, at face value,
the HDO/H$_2$O ratio ranges from 2 \% (L1527) to 22 \% (IRAS04302).
Except in the case of L1527, the ratios are much higher than those obtained in the previous
observations and theoretical works: HDO/H$_2$O $\le$ 3 \% (Dartois et al. \cite{dartois03}, Parise et al.
\cite{parise03}, Parise et al. \cite{parise05}, Aikawa et al. \cite{aikawa05}).

%L1527: 2.65E+16 ...0.5percent
%L1041: 7.938E+16...2.0
%IRAS04302: 1.323E+17...5.4
%HV Tau: 2.64E+16...4.8

\begin{figure}
\centering
\includegraphics[width=9cm]{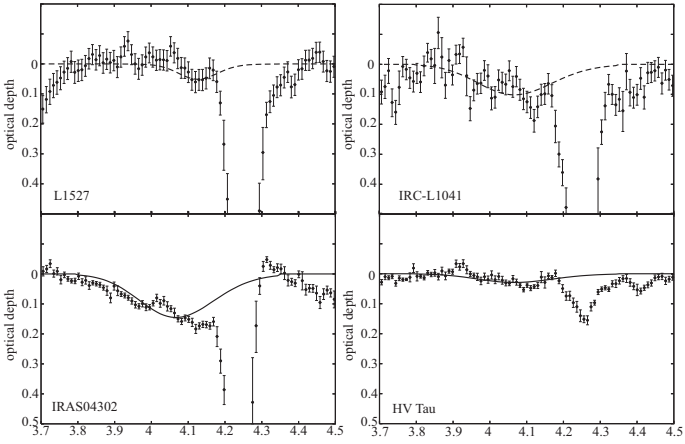}
   \caption{Spectra around the HDO absorption band. The solid line is the Gaussian with FWHM of 0.2 $\mu$m
                approximating the amorphous HDO ice feature, while the dashed line approximates the annealed HDO
                feature.}
      \label{HDO}
\end{figure}

\subsection{CO$_2$ ice}
The CO$_2$ band at 4.27 $\mu$m is clearly detected towards our targets, except ASR41 and
2MASS J1628137. To subtract the H$_2$O band at 4.5 $\mu$m, we divide the
normalized spectrum by the best-fit H$_2$O profile (red line in Fig \ref{norm_H2O}).
The absorption at 4.27 $\mu$m is then fitted by the laboratory CO$_2$ profile
convolved by the instrumental line profile (N3 profile for L1527) (solid line in Fig \ref{CO2}). To
illustrate how the profile is broadened by the instrumental line profile, the laboratory profile
before convolution is shown in the panel of IRAS04302 with the dashed line. 
While we adopted the pure CO$_2$ profile measured by Gerakines
et al. (\cite{gerakines1995}) at 10 K, the actual shape of the CO$_2$ profile depends on various factors,
such as its mixing with other ice species, grain shape, and annealing (Ehrenfreund et al \cite{ehrenfreund1997},
van Broekhuizen et al. \cite{vanbroekhuizen06}). 
We tried to fit the absorption bands with those of mixed ices of CO$_2$:CO=1:1 at 15 K, and H$_2$O:CO$_2$=
100:14 at 10 K, but the fits were worse than the pure CO$_2$ ice of 10 K.
Although the observed profile might be better fitted by other mixtures and temperatures, we do not pursue
this fitting, because the spectral resolution of the IRC is not high enough. 

In regions close to the YSO, CO$_2$ sublimates at $\sim 70$ K. We note that CO$_2$ gas has absorption lines
in almost the same wavelength range as CO$_2$ ice. Although the observed feature might be a combination
of CO$_2$ ice and gaseous CO$_2$, the low spectral resolving power makes it difficult to distinguish between gas and ice.
We tried to fit the observed feature with CO$_2$ gas of constant temperature (20K,
70K, and 150K), but the observed features are better fitted by the pure CO$_2$ ice.

We obtained the CO$_2$ ice column densities listed in Table \ref{column}. The errors listed in Table
\ref{column} were derived from the errors in the absorption area. 
For class 0-I sources, the convolution function (the instrumental line profile or the N3 profile)
might add errors to the column densities. For IRC-L1041, if we were to adopt the
N3 profile, the CO$_2$ column density should be larger by $\sim 30 \%$ than
listed in Table \ref{column}; the wing region is reproduced better, but the agreement in the peak
region is worse than in Fig \ref{CO2}. For IRAS04302, convolution with its N3 profile makes the absorption peak more
flat and thus in closer agreement with the observation, while the CO$_2$ column density is little changed.

We also detected CO$_2$ absorption towards HV Tau and HK Tau. The absorption could, at least partially,
originate in the circumstellar disks. The H$_2$O absorption towards HK Tau could originate solely in the
disk of HK Tau B (\S 4.2). 
If so, the detected CO$_2$ ice towards HK Tau should also be in the edge-on disk, since
it is unlikely that the foreground clouds contain CO$_2$ but no H$_2$O.
The CO$_2$ absorptions of HK Tau and HV Tau are, however, deeper
than 10 mJy, while the fluxes of HV Tau C and HK Tau B seem to reach a maximum in the K band, and are smaller than
10 mJy at 3.8 $\mu$m and probably at longer wavelength (\S 4.2). Hence, we derived the CO$_2$ ice column density
by normalizing the spectrum with the continuum plotted in Fig. \ref{spec}, as if the CO$_2$ ice originates in the foreground
clouds of the primary.

The relation between
column densities of CO$_2$ ice and H$_2$O ice is plotted in Figure \ref{CO2_col}. Our class
0-I stars show similar CO$_2$/H$_2$O ratios to those observed by Pontoppidan et al.
(\cite{pontoppidan2008}) towards embedded low-mass YSOs using {\it SST}.
%It is interesting that L1527 shows a relatively low
%CO$_2$/H$_2$O ratio compared with other low-mass YSOs.
%If CO$_2$ is mainly formed from CO ice, the low CO$_2$/H$_2$O ratio may support the
%scenario that L1527 is formed by a faster collapse and thus has less CO ice than other
%objects.
Towards HV Tau, HK Tau, and UY Aur, on the other hand, the column densities of H$_2$O ice and CO$_2$
ice are much lower than observed by Pontoppidan et al. (\cite{pontoppidan2008}), assuming
that the detected absorption features all originate in the foreground clouds. The CO$_2$/H$_2$O ratios toward these
class II objects, however, agree with the range of ratios for low-mass YSOs from Pontoppidan et al. (\cite{pontoppidan2008}).

We note that the $^{13}$CO$_2$ band at 4.38 $\mu$m is apparent in the spectra
of L1527 and IRC-L1041-2. The laboratory CO$_2$ profile used to fit the 4.27 $\mu$m
feature contains $^{13}$CO$_2$ with the terrestrial ratio $^{12}$C/$^{13}$C=89. 
In L1527, we fitted the 4.38 $\mu$m feature (thin solid line in Figure \ref{CO2})
with the laboratory spectrum. The absorption area ($\int \tau d\nu$) of the thin solid line is
about two times larger than that of the thick solid line, indicating that the $^{12}$CO$_2$/$^{13}$CO$_2$
ratio could be lower than the terrestrial value. However, when considering the S/N, we found that the observed
4.38 $\mu$m feature is consistent with the thick solid line in Figure \ref{CO2}, which fits the
4.27 $\mu$m feature; the terrestrial ratio $^{12}$C/$^{13}$C=89 are not significantly inconsistent with
our observational data.

\begin{figure}
\centering
\includegraphics[width=9cm]{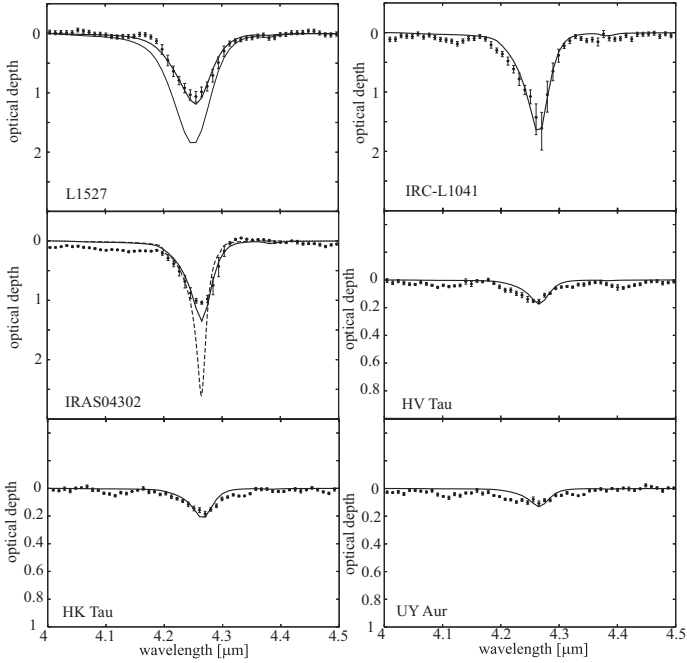}
   \caption{Spectra fitted by the convolved laboratory spectrum of CO$_2$ ice (solid line). The dashed lines
                depict the CO$_2$ feature before convolution. In the panel of L1527, the 4.38 $\mu$m feature is fitted
                by the convolved laboratory spectrum of $^{13}$CO$_2$ (thin solid line).}
      \label{CO2}
\end{figure}

\begin{figure}
\centering
\includegraphics[width=9cm]{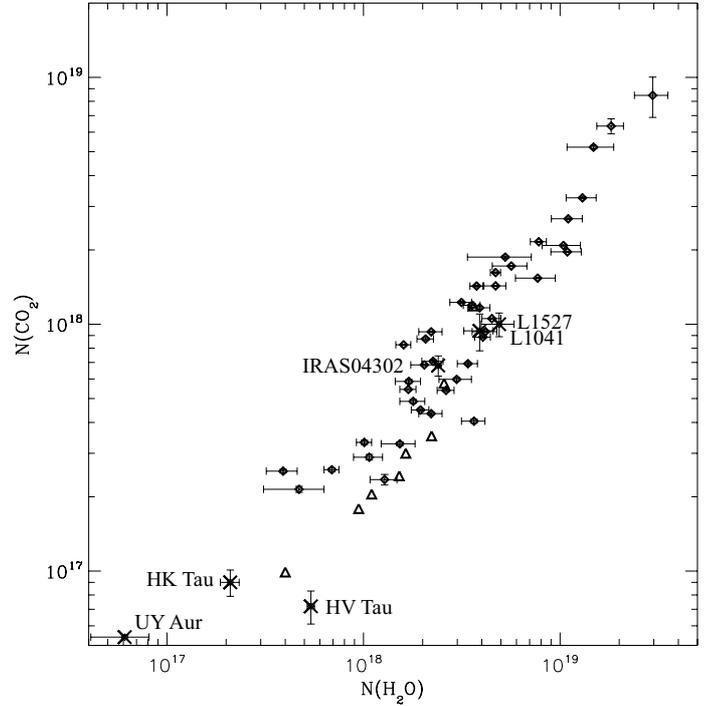}
   \caption{The relation between column densities of H$_2$O ice and CO$_2$ ice towards our
   targets (crosses) plotted over Figure 6 of Pontoppidan et al.
   (\cite{pontoppidan2008}); the diamonds are low-mass stars observed by Pontoppidan et al.
   (\cite{pontoppidan2008}), and the open triangles are the background stars from Whittet et al.
    (\cite{whittet2007}) and Knez et al. (\cite{knez05}). }
      \label{CO2_col}
\end{figure}

\subsection{CO and XCN}
We found that CO ice has an absorption band at 4.67 $\mu$m. Figure \ref{CO} shows the spectra
around this wavelength towards L1527 and IRC-L1041-2, IRAS04302, HK Tau, and UY Aur.

Both L1527 and IRC-L1041-2 clearly have double-peaked profiles. The
blue peak coincides with the XCN band, and the red peak matches the CO ice band.
Although the double peak can be reproduced by combining XCN and CO ice, the observed
features have an additional component at $\sim 4.7 -4.75$ $\mu$m. We assume that this corresponds to CO gas
because, owing to the low sublimation temperature ($\sim 20$ K), CO can be more easily sublimated
and excited near the protostar than CO$_2$.
With the low spectral resolution of IRC, gaseous CO vibrational bands cannot be observed
as lines but become a broad double-peaked ($P$ branch and $R$ branch) feature.
It is thus difficult to derive the CO gas column density from our data. However, it is important
to subtract the CO gas absorption, since the $R$ branch of the CO gas feature coincides
with the XCN feature. The distance between the $P$ branch and $R$ branch increases
with temperature. We fitted the absorption at $4.7-4.75$ $\mu$m with a CO gas profile
at 70 K that had been smoothed to the low spectral resolution (dashed lines in Figure \ref{CO}). 

After the subtraction of CO gas absorption, we fitted the XCN feature, assuming a Gaussian
peaking at 4.62 $\mu$m with FWHM of 29.1 cm$^{-1}$ (Gibb et al. \cite{gibb04})
(dotted lines in Figure \ref{CO}), and the CO ice feature using the CDE grain model
of Ehrenfreund et al. (\cite{ehrenfreund1997}) (thin solid lines in Figure
\ref{CO}). The dot-dashed line in Figure \ref{CO} depicts the laboratory spectrum before convolution,
and the thick solid lines in Figure \ref{CO} shows the sum of the three fitted components.

The absorption feature around 4.67 $\mu$m is also observed towards HK Tau and UY Aur, but
the feature does not peak at 4.67 $\mu$m. We tentatively assigned the absorption to
gaseous CO at 70 K and 150 K for HK Tau and UY Aur, respectively (thin solid lines in
Figure \ref{CO}). 

Table \ref{column} lists the estimated column densities of CO ice.
The correlation between the column densities of CO ice and H$_2$O ice is plotted in Figure \ref{CO_H2O}
over the correlation towards field stars of Whittet et al. (\cite{whittet2007}). The gray square
depicts the ice column density ratio towards class I protostar Elias 29 (Boogert \cite{boogert2000}).
Our targets show higher CO/H$_2$O ratios than Elias 29, probably because our targets are
edge-on objects. IRAS04302 shows a lower ratio of CO/H$_2$O than field stars, indicating that its
circumstellar material is warmer and/or less dense than typical molecular cloud gas.
%On the other hand, our ratios are lower than those towards field stars, except L1041. 
%CO could be sublimated to the gas phase, but adding the estimated column density of
%CO gas to CO ice does not change the CO/H$_2$O ratio significantly. Although our
%spectral resolution is not high enough to resolve CO gas feature, the absorption at
%$4.7-4.75$ $\mu$m should be deeper if we have a significant amount of CO gas.

Assuming that the XCN is OCN$^-$, we adopt the band strength of $A$(XCN)$=5 \times 10^{-17}$
cm mol$^{-1}$, and estimate the OCN$^-$ column density to be $(0.5-2.5) \times 10^{17}$ cm$^{-2}$
towards class 0-I objects (Table \ref{column}).
The column density ratio of OCN$^-$/H$_2$O ranges
from 2.2 \% (IRAS04302) to 6.4 \% (L1041-2).
Weintraub et al.  (\cite{weintraub94}) detected the XCN feature towards a T Tauri star RNO 91
and derived an upper limit to the XCN/H$_2$O ratio of $\lesssim 9$ \%.  More recent works, on the other
hand, have derived much lower XCN/H$_2$O ratio towards low-mass YSOs.
van Broekhuizen et al.
(\cite{vanbroekhuizen05}) observed low-mass YSOs with VLT; the column density ratio of OCN$^-$/H$_2$O is
$\le 0.85$ \%. Although CO gas can contribute to the absorption at $\sim 4.62$ $\mu$m in our spectra,
its contribution is subtracted by referring to the absorption at $4.7-4.75$ $\mu$m.
In laboratory experiments, thermal or UV processing of N-bearing ice produces OCN$^-$ (Schutte \&
Greenberg \cite{schutte97}).
Since our targets are edge-on objects, our lines of sight may preferentially trace the surface region
of the disks (or the dense torus region of the envelope) where OCN$^{-}$ is produced by UV irradiation.
The OCN$^-$ column densities towards IRC-L1041 and IRAS04302, however, should be taken with caution,
since there is residual absorption in the $4.5-4.6$ $\mu$m region (Fig \ref{CO}). If there are other unknown carriers
at these wavelengths, the estimated OCN$^-$ column densities are upper limits.

Both L1527 and IRC-L1041-2 show weak absorption at $\sim 4.78$ $\mu$m. Although it is
close to the peak of $^{13}$CO ice feature, the observed peak is at a slightly shorter wavelength
than that of $^{13}$CO. 

\begin{figure}
\centering
\includegraphics[width=9cm]{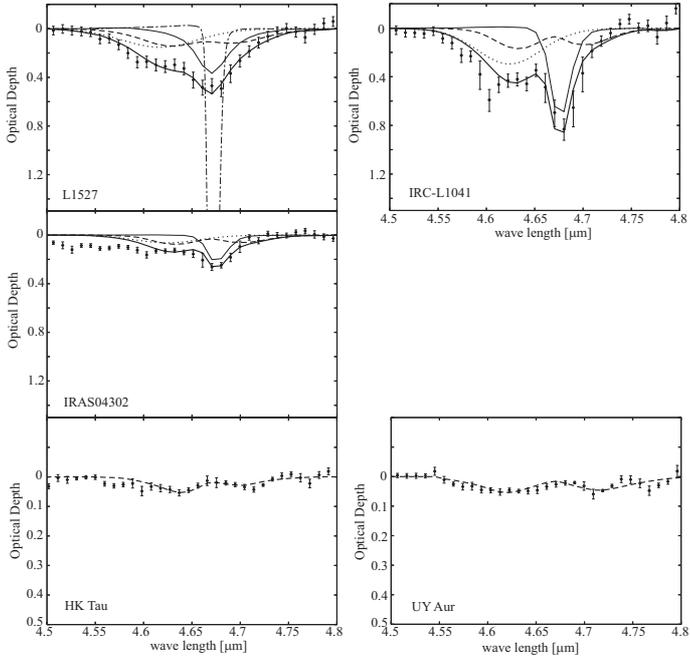}
  \caption{The spectra around the CO ice absorption band fitted by a combination (thick solid line) of CO gas
(dashed line),   XCN feature (dotted line), and CO ice on the CDE grain model (thin solid line). The dot-dashed line in the
panel of L1527 depicts the CO ice feature before the convolution with the instrumental line profile.}
      \label{CO}
\end{figure}

\begin{figure}
\centering
\includegraphics[width=9cm]{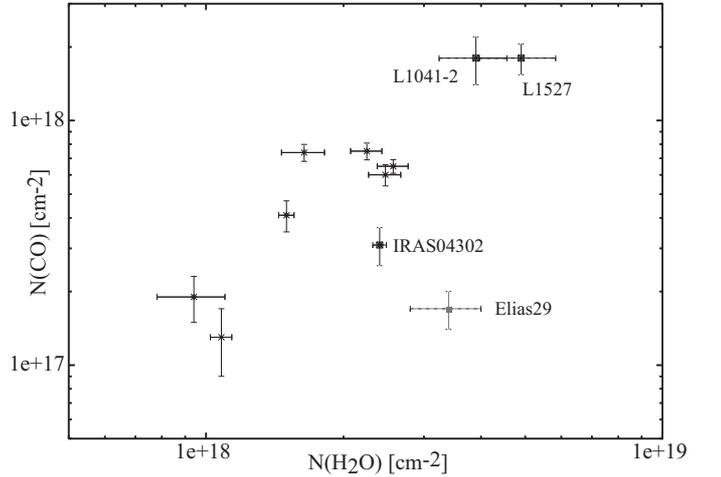}
   \caption{The relation between the column densities of H$_2$O ice and CO ice towards our
   targets (square). The crosses are ice columns towards field stars  observed by Whittet et al.
   (\cite{whittet2007}). The gray square depicts the ice column density towards the low-mass
   class I protostar Elias29 (Boogert et al. \cite{boogert2000}).}
      \label{CO_H2O}
\end{figure}

%KOKOKARA

\subsection{OCS}
IRC-L1041-2 has an absorption feature at $\sim 4.9$ $\mu$m. This coincides with the OCS ice band,
which can be approximated by a Gaussian peaking at 4.91$\mu$m with a FWHM of 19.6 cm$^{-1}$
(Gibb et al. \cite{gibb04}; Palumbo, Geballe \& Tielens \cite{palumbo97}; Hudgins et al. \cite{hudgins1993}).
Although the feature is at the edge of the
observable wavelength range of the IRC and should be treated with caution, it does not disappear even if
we change the combination of frames to be stacked. We fitted the spectrum with a Gaussian
(Figure \ref{OCS}). Assuming a band strength of $1.7 \times 10^{-18}$ cm molecule$^{-1}$
(Hudgins et al. \cite{hudgins1993}), the OCS column
density towards IRC-L1041-2 is $6.2 \times 10^{16}$ cm$^{-2}$. If the abundance of H$_2$O ice
relative to hydrogen nuclei is $1\times 10^{-4}$, the hydrogen column density in the line of sight is
$3.9\times 10^{22}$ cm$^{-2}$, and the relative abundance of OCS ice to hydrogen is $1.6 \times 10^{-6}$,
which is much higher than the abundance of gaseous OCS, $1\times 10^{-9}$, observed towards TMC-1
(Irvine et al. \cite{irvine1987}).

\begin{figure}
\centering
\includegraphics[width=6.5cm]{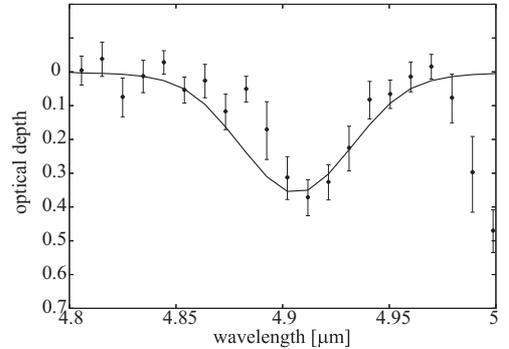}
   \caption{Spectrum of IRC-L1041-2 at 4.9 $\mu$m. The solid line is a Gaussian with FWHM 19.6 cm$^{-1}$
approximating the OCS absorption feature.}
      \label{OCS}
\end{figure}

\section{Discussion}
\subsection{Location of the Ice}

Young stellar objects have a complicated structure: a protostar, circumstellar disk, envelope, and outflow cavity.
Furthermore, some objects might be accompanied by either ambient or foreground clouds.
In our observation of edge-on YSOs, the light source is not necessarily the central star. The NIR
light of $3-5$ $\mu$m can originate from, for example, a wall of the outflow cavity that scatters
the stellar light. The light source can thus be spatially extended relative to the projected spatial scale
of ice abundance structures, even if it looks like a point source
(e.g. Tobin et al. \cite{tobin08}; \cite{tobin10}), and the absorbing material in the envelope and/or
disk also has non-uniform spatial distributions.
Although the radiative transfer modeling of such an envelope and disk system is beyond the scope of
the present work, it is important to discuss the location of ice in our targets.

\subsubsection{Class 0-I objects}
Among the three class 0-I objects, L1527 is the most well-studied. Its bolometric temperature is less than 70 K,
suggesting that the object is deeply embedded in dense gas (Furlan et al. \cite{furlan08}).
Tobin et al. (\cite{tobin08}) constructed an
axisymmetric model of the envelope and disk with an outflow cavity referring to the {\it SST} images and SED
of the object. According to the model, the "central object" we see in the N3 image is scattered light from the
"inner cavity" with a size of $\sim 100$ AU. If so, the ice absorption bands in our spectrum most likely originate in the
envelope rather than the disk. The envelope density ranges from about $10^8$ cm$^{-3}$ to $10^5$ cm$^{-3}$
at $\sim 100$ AU to 15000 AU. The integrated envelope column density in the line of sight is $\sim 1$
g cm$^{-2}$, which corresponds to a hydrogen column density of $N_{\rm H} \sim 4 \times 10^{23}$ cm$^{-2}$.
This would, however, be an overestimate, because the envelope structure is not axisymmetric
in reality. 
%Unlike other ice absorptions, CO gas and OCN$^-$ features would
%arise from the region near the cavity, where the temperature is warm and UV radiation is intense.

Owing to the similarity of the continuum spectra, we expect L1041 and L1527 to have a similar physical
structure. Varying depths of ice absorptions among these objects would thus probe the difference in the
ice composition in the envelope. 
%It is interesting that the relative ice abundances of CO and CO$_2$ to H$_2$O
%are lower towards L1527 than L1041. It would support the idea that L1527 is formed by a faster collapse than typical
%low-mass YSOs. Alternatively, the high CO ice column density towards L1041 indicates that it is deeply embedded
%in clouds.

The SED modeling of the YSOs suggests that the envelope of IRAS04302 is slightly less dense than L1527
(Furlan et al. \cite{furlan08}). We found smaller column densities of ices and lower CO/CO$_2$ ice ratios
towards IRAS04302 than L1527, which also indicates that IRAS04302 is more evolved and has a less massive envelope.
%The CO$_2$ ice absorption, however, is more complicated. We notice that the 15.2 $\mu$m CO$_2$ ice feature
%is much less significant than that of L1527 (Furlan \cite{furlan08}), while the column density of CO$_2$ ice derived
%from our observation is similar to that of L1527.
%Pontoppidan et al. (\cite{crbr05}) calculated the infrared spectra of protoplanetary disk models
%with ices; they showed that the ratio of optical depth at 4.27 $\mu$m and 15.2 $\mu$m
%changes significantly with the inclination angle. Although the model of Pontoppidan et al. (\cite{crbr05}) does not
%include the envelope, a similar dependence on inclination might be possible in torus envelopes.

Although our observations probe ice in envelopes rather than disks, the envelope material along our lines of sight
(perpendicular to the rotation axis and outflow) is more likely to end up in the forming disks than be dissipated in the
clouds. The high CO/CO$_2$ ice ratio indicates that the majority of gases in the lines of sight are still dense and cold.
The CO$_2$/H$_2$O ratio towards our objects is similar to the ratio observed by Pontoppidan et al.
(\cite{pontoppidan2008}), who observed low-mass YSOs with various inclinations; the CO$_2$/H$_2$O ratio does not
seem to vary with the inclination angle.

The CO gas absorption and deep absorption feature of OCN$^-$, on the other hand, would originate in
the warm ($T\ge 20$ K) and irradiated region near the protostar. However, other signs of heating
by the protostar, such as crystalline feature of H$_2$O ice (Schegerer \& Wolf \cite{schegerer10})
and CO$_2$ gas, are not found in our data. We may need higher spatial resolution
and spectral resolution to detect these features.

\subsubsection{Edge-on class II objects: ASR41 and 2MASS J1628137}

We estimated the H$_2$O ice column density from our observations to be $7.8 \times 10^{17}$ cm$^{-2}$
towards ASR41 and $6.7\times 10^{17}$ cm$^{-2}$ towards 2MASS J1628137. These values are
apparently much lower than we expect from the theoretical prediction of high H$_2$O ice
abundance n(H$_2$O)/n$_{\rm H}$ $\sim 10^{-4}$ in the disk midplane. The shallow absorption is, however,
caused by a geometric effect. Pontoppidan et al. (\cite{crbr05}) predicts that the 3 $\mu$m water band
is the deepest for the disk inclination of $\sim 70^{\circ}$ and becomes shallower for higher inclination
angles. If the inclination angle is $\ge 72^{\circ}$, the optical depth of the band is predicted to be $\sim 0.6$,
which is comparable to our observation of ASR41. Since the light source (scattered stellar light) is more extended
than the absorbing material, Eq, (2) underestimates the ice column density towards edge-on disks.

Possible contributions to the observed ice column densities from foreground clouds should also be considered. 
In the Taurus  molecular cloud, the average H$_2$O abundance
is about $8.6\times 10^{-5}$ relative to hydrogen nuclei at $A_{\rm v} \ge 3$ mag (Whittet et al. \cite{whittet93};
Chair et al. \cite{chair11}).
The H$_2$O ice is not detected at lower $A_{\rm v}$, which indicates that H$_2$O ice is easily destroyed by \
photolysis or photodesorption near the cloud surface.
Hodapp et al. (\cite{hodapp04})
estimated that the density and size of the molecular cloud around ASR 41 are $2.0 \times 10^{-20}$ g cm$^{-3}$
and $10^4$ AU, which corresponds to $A_{\rm V}\sim 1$ mag. Although the relative abundance and the threshold
$A_{\rm v}$ varies from cloud to cloud, the column density of the foreground gas towards ASR41,
$A_{\rm v}\sim 1$ mag, is too small to contribute significantly to the H$_2$O ice absorption band.
Therefore, the observed H$_2$O ice most likely originates in the disk around ASR41.

The foreground visual extinction towards 2MASSJ1628137, on the other hand, is estimated to be
$A_{\rm V}=2.1\pm 2.6$ mag (Grosso et al. \cite{grosso03} ). Assuming that the H$_2$O ice abundance and threshold
$A_{\rm v}$ are the same as in Taurus, the upper limit of the foreground H$_2$O ice column density is
$\sim 2 \times 10^{17}$ cm$^{-2}$, which is smaller than the observed H$_2$O ice column ($6.7 \times 10^{17}$
cm$^{-2}$). Hence, we conclude that most or all of the H$_2$O ice observed towards 2MASSJ1628137 originates in the disk.

\subsubsection{Other class II objects}

HV Tau and HK Tau are multiple systems. As the separation between the multiple components
in these systems is smaller than the size of our PSF in the NIR, we cannot extract the spectra of the
faint edge-on objects.
Terada et al. (\cite{terada07}) observed $2-2.5$ $\mu$m and $3-4$ $\mu$m spectra of these edge-on
objects using Subaru and detected deep water absorption ($\tau\sim 1-1.5$). The continuum flux of HV Tau C
is $\sim 7.75$ mJy at 2.2 $\mu$m and $\sim 7.14$ mJy at 3.8 $\mu$m, while the flux of HK Tau B
is $\sim 15.9$ mJy at 2.2 $\mu$m and $\sim 9.85$ mJy at 3.8 $\mu$m.
Apparently, our spectra are dominated by the bright primaries. Both H$_2$O and CO$_2$ ice absorptions towards
HV Tau should originate in both the disk and foreground clouds, since the depths of the absorption bands are
larger than the continuum flux of the HV Tau C.
Towards HK Tau, the depth of the H$_2$O absorption is comparable to the continuum flux of HK Tau B;
if the H$_2$O absorption originates in the disk around HK Tau B, the H$_2$O ice column density is $\sim 3 \times 10^{18}$
cm$^{-2}$, which is comparable to the value obtained by Terada et al. (\cite{terada07}). 
The CO$_2$ absorption, on the other hand, is deeper than the continuum flux of HK Tau B, and thus should 
originate at least partially in the foreground clouds.
We also detected shallow H$_2$O ice bands towards UY Aur, which again originate in the foreground component,
since the inclination of the disk is $\sim 42$ degrees. 
%These foreground clouds are rare examples showing CO$_2$ ice absorption
%but no CO ice absorption. They must be warmer than the CO sublimation temperature ($\sim 20$ K).

If the H$_2$O absorptions towards HV Tau and HK Tau originated in the foreground clouds,
the estimated H$_2$O ice column densities would be smaller than the H$_2$O ice column densities in the edge-on disks
obtained by Terada et al. (\cite{terada07}) by about an order of magnitude; the H$_2$O ice column density would be
$3.31 \times 10^{18}$ cm$^{-2}$ for HK Tau B, and $(2.69-4.20) \times 10^{18}$ cm$^{-2}$ for HV Tau C. 
Our observations thus support the argument of Terada et al (\cite{terada07})
that the disk component overwhelms the foreground component towards HK Tau B and HV Tau C.

\section{Conclusions}
We have observed ice absorption bands at $2.5-5$ $\mu$m towards eight low-mass YSOs: three class 0-I
protostellar cores with edge-on geometry, two edge-on class II objects, two multiple systems with edge-on
class II, and one not-edge-on class II object.

Towards the class 0-I objects, L1527, IRC-L1041-2, and IRAS04302, we have detected abundant H$_2$O, CO$_2$,
and CO ice in the envelope.
The column density ratio of CO$_2$ to H$_2$O ice is $21-28$ \%, which coincides with the ratio
observed by {\it SST} towards YSOs with various inclinations.
The weak absorption at $\sim 4.1$ $\mu$m can be fitted by HDO ice;
the HDO/H$_2$O ratio ranges from 2 \% to 22 \%.
The absorption in the vicinity of the CO band (4.76 $\mu$m) is double-peaked
and fitted by combining CO ice, OCN$^-$, and CO gas. The large column density of CO ice suggests that the
envelope is still very dense and cold, while OCN$^-$ and CO gas features would originate in the region close
to the protostar. The column density of OCN$^-$ is as high as
$2-6$ \% relative to H$_2$O, which is much higher than previous observations. Our lines of sight (high inclinations
from the rotation axis) may preferentially trace the regions with high UV irradiation, such as the surface of a forming
disk and/or torus envelope.

The spectrum of IRAS04302 includes the 3.5 $\mu$m absorption band, but the feature does not match either CH$_3$OH or CH$_4$.
%The WCCC object, L1527, does not show either the CH$_3$OH feature or the CH$_4$ feature
%at 3.5 $\mu$m, and has relatively low ice column density ratios of CO$_2$/H$_2$O and CO/H$_2$O,
%which could be consistent with the idea that WCCCs have experienced a fast collapse resulting in relatively low
%abundances of ice (Sakai et al. \cite{nami08}).
An OCS absorption band is tentatively detected towards IRC-L1041-2.
 
Towards the edge-on class II objects, ASR41 and 2MASS J1628137-243139, we have detected the H$_2$O band.
The low optical depth of the water feature is due to geometrical effects (Pontoppidan et al. \cite{crbr05}).
The detected water ice mainly originates in the disk.

Both HK Tau B and HV Tau C are edge-on class II objects in multiple systems. In our spectra, which are dominated by
the primaries, we have detected the absorption of H$_2$O ice and CO$_2$ ice. Ices in both the edge-on disks and
foreground clouds would contribute to the absorption, although H$_2$O absorption towards HK tau could originate
solely in the disk. Even if the observed features are due to the foreground clouds, the H$_2$O ice column densities
in the clouds are much smaller than those observed towards HK Tau B and HV Tau C with Subaru
(Terada et al. \cite{terada07}), which confirms that the ice columns of the latter must originate in disks. 
The foreground H$_2$O ice and CO$_2$ ice are also detected towards UY Aur, which is not an edge-on system.
We have tentatively detected CO gas towards HK Tau and UY Aur.

\begin{acknowledgements}
We would like to thank the project members of AKARI and AFSAS team for their help with the
observation. We are grateful to Dr. D. Heinzeller for providing models of gas absorption features
and to Prof. Whittet for his permission to use the CO/H$_2$O plot from Whittet et al. (2007).
We would like to thank the anonymous referee for helpful comments that helped to improve the manuscript.
This work was supported by a grant-in-aid for scientific research (19740103, 23540266, 23103004).
\end{acknowledgements}


\begin{thebibliography}{}

   \bibitem[2005]{aikawa05} 
       Aikawa, Y., Herbst, E., Roberts, H., Caselli, P. 2005, ApJ, 620, 330

   \bibitem[2008]{aikawa08} 
       Aikawa, Y., Wakelam, V., Garrod, T.R. \& Herbst, E. (2008) ApJ, 674, 984

%   \bibitem[2010]{bourke}
%      Bourke, T.

   \bibitem[2000]{AWB2000}
      Andr\'{e}, P., Ward-Thompson, D. \& Barsony, M. (2000) Protostars and Planets IV, 59

    \bibitem[2000]{boogert2000}
    Boogert, A.C.A., Tielens, A.G.G.M., Ceccarelli, C. et al.
    2000, A\&A, 360, 683

     \bibitem[2002]{boogert2002}
     Boogert, A.C.A., Blake, G.A., \& Tielens, A.G.G.M. 2002, ApJ, 577, 271

     \bibitem[2004]{boogert04}
     Boogert, A.C.A., Blake, G.A. \&  {\"O}berg, K. 2004, ApJ, 615, 344

     \bibitem[2008]{boogert08}
	Boogert, A.C.A., Pontoppidan, K.M., Knez, C. et al.
    2008, ApJ, 678, 985
    
    \bibitem[2011]{boogert11}
    Boogert, A.C.A., Huard, T.L., Cook, A.M. et al.
    2011, ApJ, 729, 92 

    \bibitem[1996]{brooke1996}
    Brooke, T.Y., Sellgren, K. \& Smith, R.G. 1996, ApJ, 459, 2009
    
    \bibitem[2011]{chair11}
    Chair, J.E., Pendleton, Y.J., Allamandola, L.J. et al. 2011, ApJ, 731, 9

    \bibitem[2003]{dartois03}
    Dartois, E.,  Thi, W.F.,  Geballe, T.R. et al. 2003, A\&A, 399, 1020

   \bibitem[1997]{ehrenfreund1997}
      Ehrenfreund, P., Boogert, A.C.A., Gerakines, P.A., Tielens, A.G.G.M.,
      van Dishoeck, E.F. 1997, A\&A 328, 649

   \bibitem[2008]{furlan08} 
Furlan, E., McClure, M.,Calvet, N. et al. 2008,
ApJS, 176, 184

   \bibitem[2007]{geers07}
      Geers, V.C., van Dishoeck, E.F., Visser, R. et al. 2007, A\&A 476, 279

   \bibitem[1995]{gerakines1995}
      Gerakines, P.A., Schutte, W.A., Greenberg, J.M., van Dishoeck, E.F. 1995,
      A\&A. 296, 810

   \bibitem[1996]{gerakines1996}
      Gerakines, P.A., Schutte, W.A., Ehrenfreund, P. 1996, A\&A, 312, 289

%   \bibitem[1999]{gerakines1999}
%      Gerakines, P.A. et al.(1999), ApJ, 522, 357

    \bibitem[2000]{Gibb2000}
   Gibb, E. L., Whittet, D. C. B., Schutte, W. A., et al. 2000, ApJ, 536, 347

   \bibitem[2004]{gibb04} Gibb, E.L., Whittet, D.C.B., Boogert, A.C.A.,
        \& Tielens, A.G.G.M. 2004, ApJS, 151, 35

   \bibitem[1991]{grim1991}
   Grim, R.J.A, Baas, F., Geballe, T.R., Greenberg, G.M. \& Schutte, W. 1991, A\&A 243, 473

   \bibitem[2003]{grosso03}
      Grosso,N., Alves, J., Wood, K. et al.
      2003, ApJ, 586, 296-305

   \bibitem[2007]{hioki07}
      Hioki, T.,  Itoh, Y.,  Oasa, Y. et al.
      2007, ApJ, 134, 880

   \bibitem[2004]{hodapp04}
      Hodapp, K.W., Walker, C.H., Reipurth, B. et al. 2004, ApJL, 601, L79

   \bibitem[2009]{honda09} 
       Honda, M., Inoue, A.K., Fukagawa, M. et al.
       ApJL, 690, 110

    \bibitem[1993]{hudgins1993}
    Hudgins, D. M., Sandford, S. A., Allamandola, L. J., \& Tielens, A. G. G. M. 1993, ApJS, 86, 713

    \bibitem[1987]{irvine1987}
    Irvine, W.M., Goldsmith, P.F. \& Hjalmarson, A. 1987, in interstellar processes,
    eds. Hollenbach, D.J. \& H.A.Thronson Jr., p561

    \bibitem[2005]{knez05}
    Knez, C., Boogert, A. C. A., Pontoppidan, K. M. et al. 2005, ApJ, 635, L145

    \bibitem[1991]{lacy91}
    Lacy, J.H., Carr, J.S., Evans, N.J.II et al. 1991, ApJ, 376, 556

    \bibitem[1983]{leger1983}
    L\'{e}ger, A., Gauthier, S., Defourneau, D., \& Rouan, D. 1983, A\&A, 117, 164

   \bibitem[2000]{monin2000}
   Monin, J.-L. \& Bouvier, J. 2000, A\&A, 356, L75

   \bibitem[2007]{murakami07}
   Murakami, H., Baba, H., Barthel, P. et al. 2007, PASJ, 59, 369

   \bibitem[2000]{murakawa00}
     Murakawa, K., Tamura, M., \& Nagata, T. 2000, ApJS, 128, 603

   \bibitem[2008]{oberg_ch4}
   {\"O}berg, K.I., Boogert, A.C.A., Pontoppidan, K.M. et al. 2008, ApJ, 678, 1032

   \bibitem[2007]{ohyama07}
      Ohyama, Y., Onaka, T., Matsuhara, H. et al. 2007, PASJ, 59, 411

   \bibitem[2007]{onaka07}
      Onaka, T., Matsuhara, H., Wada, T. et al. 2007, PASJ, 59, 401

    \bibitem[1997]{palumbo97}
    Palumbo, M.E., Geballe, T.R., \& Tielens, A.G.G.M. 1997, ApJ, 479, 839

    \bibitem[2003]{parise03}
    Parise, B.,  Simon,T.,  Caux, E. et al. 2003, A\&A, 410, 897

    \bibitem[2005]{parise05}
    Parise, B., Caux, E., Castets, A. et al. 2005, A\&A, 431, 547

   \bibitem[2003]{pontoppidan2003}
      Pontoppidan, K. M., Fraser, H. J., Dartois, E. et al. 2003, A\&A, 408, 981

   \bibitem[2003]{ponto_ch3oh}
     Pontoppidan, K.M., Dartois, E., van Dishoeck, E.F., Thi, W.-F., \& d|Hendecourt, L.2003, 
     A\&A, 404, L17
      
   \bibitem[2005]{crbr05}
       Pontoppidan, K.M. \& Dullemond C.P. 2005, A\&A, 435, 595

    \bibitem[2008]{pontoppidan2008}
    Pontoppidan,K.M., Boogert, A.C.C., Fraser,H.J. et al. 2008, ApJ, 678, 1005

   \bibitem[2008]{nami08} 
       Sakai,N., Sakai, T., Hirota, T.\& Yamamoto, S. (2008) ApJ, 672, 371

   \bibitem[2007]{sakon}
     Sakon, I., Onaka, T., Wada, T. et al. 2007, PASJ, 59, 483
  
   \bibitem[1997]{schutte97}
     Schutte, W.A. \& Greenberg, J.M., 1997, A\&A, 317, L43

   \bibitem[2010]{schegerer10}
   Schegerer, A.A. \& Wolf, S. 2010, A\&A, 517, 87

   \bibitem[1998]{simon98}
      Simon, M., Holfeltz, S.T. \& Taff, L.G. 1998, ApJ, 469, 890

    \bibitem[1989]{smith1989}
    Smith, R. G., Sellgren, K., \& Tokunaga, A. T. 1989, ApJ, 344, 413 
    
    \bibitem[1998]{stapelfeldt98}
    Stapelfeldt, K.R., Krist, J.E., Menard, F. et al. 1998
    ApJL, 502, 65

   \bibitem[2007]{terada07} 
       Terada, H., Tokunaga, A. T., Kobayashi, N. et al. 2007, ApJ, 667, 303
   
   \bibitem[1999]{teixeira99}
   Teixeira, T. C., \& Emerson, J. P. 1999, A\&A, 351, 303
   
   \bibitem[2002]{thi02} 
       Thi, W.F., Pontoppidan, K.M., van Dishoeck, E.F., Dartois, E.
       \& d'Hendecourt, L. 2002, A\&A, 394, L27
       
   \bibitem[2008]{tobin08}
       Tobin, J.J., Hartmann, L., Calvet, N. \& D'Alessio, P. 2008, ApJ, 679, 1364
       
   \bibitem[2010]{tobin10}
       Tobin, J.J., Hartmann, L., Looney, L.W. \& Chiang, H.-F. 2010, ApJ, 712, 1010


  \bibitem[2004]{vanbroekhuizen04}
     van Broekhuizen, F.A., Keane, J. V., \& Schutte, W. A. 2004,
     A\&A, 415, 425

  \bibitem[2005]{vanbroekhuizen05}
     van Broekhuizen, F.A., Pontoppidan, K.M., Fraser, H.J., \& van Dishoeck, E.F.
     A\&A 441, 249

   \bibitem[2006]{vanbroekhuizen06}
      van Broekhuizen, F.A., Groot, I.M.N., Fraser, H.J.,
       van Dishoeck, E.F. \& Schlemmer, S. 2006, A\&A, 451, 273


    \bibitem[1996]{vanDishoeck1996}
    van Dishoeck, E.F.,  Helmich, F.P.,  de Graauw, Th. et al. 1996, A\&A, 315, L349


   \bibitem[2009]{visser09}
   Visser, R., van Dishoeck, E.F., Doty, S.D. \& Dullemond, C.P. 2009, A\&A 495, 881

   \bibitem[1994]{weintraub94}
   Weintraub, D.A., Tegler, S.C., Kastner, J.H., \& Rettig, T. 1994, ApJ, 423, 674

  \bibitem[1993]{whittet93} Whittet, D.C.B. 1993,
    Dust and Chemistry in Astronomy (Institute of Physics Publishing, Bristol
     and Philadelphia), 9
  
  \bibitem[1998]{whittet98}
  Whittet, D.C.B., Gerakines,P.A., Tielens, A. G. G. M. et al.
  1998, ApJ, 498, L159

    \bibitem[2007]{whittet2007}
    Whittet, D.C.B., Shenoy, S.S., Bergin, E.A. et al.
    2007, ApJ, 655, 332

   \bibitem[1998]{woitas98}
      Woitas, J. \& Leinert, Ch. 1998, A\&A, 338, 122
      
   \bibitem[2003]{wolf03}
   Wolf, S., Padgett, D., \& Stapelfeldt, K.R. 2003, ApJ, 588, 373

   \bibitem[2009]{zaowski09}
      Zasowski, G, Kemper, F., Watson, D.M. et al. 2009, ApJ, 694, 459

\end{thebibliography}
\end{document}